\documentclass[table]{article}
\usepackage{array,tabularx}
\usepackage{todonotes}

\usepackage[skip=0pt]{caption}
\PassOptionsToPackage{usernames,dvipsnames,svgnames,table}{xcolor}
\usepackage[export]{adjustbox}
\usepackage{enumerate}
\usepackage{subcaption}
\usepackage{mathtools}
\usepackage{amsmath}
\usepackage{amsthm}
\usepackage{amssymb}
\usepackage{multicol}
\usepackage{amsfonts}
\usepackage[colorlinks=false, breaklinks=true]{hyperref}
\usepackage{color}
\usepackage{booktabs}
\usepackage{paralist}
\usepackage{textcomp}
\usepackage{graphicx}
\usepackage{rotating}
\newcommand{\red}[1]{\textcolor{red}{#1}}
\providecommand{\keywords}[1]
{
  \small
  \textbf{\textit{Keywords---}} #1
}

\title{Large-scale analysis of grooming in modern social networks}

\begin{document}


\author{Nikolaos Lykousas$^{1,2}$ and Constantinos Patsakis$^{1,3}$\\
\small {\href{mailto:nlykousas@data-centric.eu}{nlykousas@data-centric.eu}, \href{mailto:kpatsak@unipi.gr}{kpatsak@unipi.gr}}\\
\small $^{1}$Department of Informatics, University of Piraeus,\\\small 80 Karaoli \& Dimitriou str., 18534 Piraeus, Greece\\
        \small $^{2}$Data-Centric,\\\small Str. Nerva Traian 3, Bucharest, Romania\\
        \small $^{3}$Information Management Systems Institute,\\\small Athena Research Center, Artemidos 6, Marousi 15125, Greece\\
}
\date{}

 \maketitle
  \begin{abstract}
      Social networks are evolving to engage their users more by providing them with more functionalities. One of the most attracting ones is streaming. Users may broadcast part of their daily lives to thousands of others world-wide and interact with them in real-time. Unfortunately, this feature is reportedly exploited for grooming. In this work, we provide the first in-depth analysis of this problem for social live streaming services. More precisely, using a dataset that we collected, we identify predatory behaviours and grooming on chats that bypassed the moderation mechanisms of the LiveMe, the service under investigation. Beyond the traditional text approaches, we also investigate the relevance of emojis in this context, as well as the user interactions through the gift mechanisms of LiveMe. Finally, our analysis indicates the possibility of grooming towards minors, showing the extent of the problem in such platforms.
  \end{abstract}


\section{Introduction}
The recent advances in telecommunications have unleashed the potentials of sharing and exchanging content, changing radically the way we interact with others online. By lifting many bandwidth barriers, users may generate and share arbitrary content and disseminate it instantly to millions of users. As a result, we see the dominance of Social Networks and Media in various aspects of our daily lives.

This radical shift and penetration of mobile devices have led millions of people and youngsters to use them on a daily basis. While most social networks have specific policies about use from minors, in practice, this policy is bypassed. Minors declare fake ages to register to service providers and end up using the services as normal users. While this might not be noticed or be overseen by service providers, this is not the case of users. Unfortunately, there are thousands of users who maliciously target minors. Of specific interest is the case of {\it grooming}. Grooming refers to the process by which an offender prepares a victim for sexually abusive behaviour. More precisely, according to Craven et al. \cite{craven2006sexual}:
\begin{quote}
\textit{[Grooming is]... a process by which a person prepares a child, significant others, and the environment for the abuse of this child. Specific goals include gaining access to the child, gaining the child's compliance, and maintaining the child's secrecy to avoid disclosure. This process serves to strengthen the offender's abusive pattern, as it may be used as a means of justifying or denying their actions...}
\end{quote}
Apparently, the case of child grooming is of extreme importance due to the impact that it can have in the children's lives. In fact, despite the measures that social networks might have already taken, they do not seem to be successful at all\footnote{\url{https://www.bbc.com/news/uk-47410520}}. To this end, it is necessary to investigate how grooming in social networks works and how groomers manage to bypass the policies and filters that are set by social networks. In terms of verbal content, currently, there is only one available dataset from the Perverted Justice website\footnote{\url{http://perverted-justice.com/}}. The organisation behind this website, Perverted Justice Foundation, Inc., has recruited volunteers to carry out sting operations in which they appear as minors to several online services and record the interactions with them. Their operations have made a tremendous positive impact as they have led to the conviction of more than 620 offenders. While undoubtedly, this is a huge contribution, the problem persists, and the provided dataset is rather old to be used for modern filters.

\noindent\textbf{Motivation.} The past few years, there is a steady increase of reports in mainstream media and officials\footnote{\url{https://www.nspcc.org.uk/what-we-do/news-opinion/3000-new-grooming-offences/}} regarding the exploitation of social networks for grooming. The problem regardless of the age factor, is rather big and stigmatises the life of thousands of people. The emergence of new social networks, allowing live streaming to potentially thousands of users along with traditional chatting and appraisal methods of traditional social networks have the potential to be further exploited for grooming.

The findings discussed in \cite{lykousas2018adult} demonstrated that the moderation systems used by the LiveMe platform at that time were highly ineffective in suspending the accounts of deviant users producing adult content. Notably, in the same year, FOX 11; a major mainstream media outlet, reported \cite{fox} that:
\begin{quote}
    \textit{A FOX 11 investigation has found that pedophiles are using the popular live streaming app LiveMe to manipulate underage girls into performing sexual acts, reward them with virtual currency, and then post screen captures or recordings of the girls online to be sold and distributed as child porn.}
\end{quote}
As such, it is reasonable to assume that the adult content problem and the sexual grooming behaviours identified by FOX 11 are related to some extent. In this work, we aim to unveil communication patterns of sexual groomers in the context of social live streaming services.

\noindent\textbf{Main contributions.}
The contributions of this work are multifold. First, we facilitate research in this field and the generation of new filters and algorithms to detect such predatory behaviour through the release of a large-scale dataset of both verbal and non-verbal interactions (e.g. likes and rewards) in a Social Live Streaming Service. Due to its nature, the dataset is available only to researchers and law enforcement agencies upon request via Zenodo\footnote{\url{https://zenodo.org/record/3560365}}. Second, we analyse the basic characteristics of the verbal content. Our analysis illustrates how such predatory behaviour bypasses the filters of service providers by, e.g. altering some ``bad words'', or by using emojis. Notably, to the best of our knowledge, this is the first work highlighting the role of emojis in grooming. Then, based on our analysis, we manage to identify chats where grooming is performed. Moreover, we analyse non-verbal interactions between users that differentiate chats where grooming is performed from the others. Finally, our analysis shows that it is possible to identify illegal actions, such as the grooming of minors.

\noindent\textbf{Organisation of the article.} The rest of this work is organised as follows. In the next section, we provide an overview of the related work on the detection of deviant behaviour and grooming. Then, we provide the legal and ethical justification of collecting such data and GDPR compliance assessment. Section \ref{sec:dataset} provides an overview of our dataset. In Section \ref{sec:analysis}, we make an analysis of our dataset and provide some insight into it. Afterwards, in Section \ref{sec:lda}, we investigate possible modelling of grooming behaviours using both verbal and non-verbal features. Finally, the article concludes summarising our contributions and discussing ideas for future work.

\section{Related work}
 Colleto et al. \cite{Coletto2017} aimed at going beyond previous studies that considered deviant groups in isolation by observing them in context. In particular, they attempted to answer questions relevant to the deviant behaviours related with pornographic material in the social media context, such as i) how much deviant groups are structurally secluded from the rest of the social network, and what are the characteristics of their subgroups who build ties with the external world; ii) how the content produced by a deviant community spreads and what is the entity of the diffusion which reaches users outside the boundaries of the deviant community who voluntarily or inadvertently access the adult content, and iii) what is the demographic composition of producers and consumers of deviant content and what is the potential risk that young boys and girls are exposed to it. Very interestingly they find that while deviant communities may have limited size, they are tightly connected and structured in subgroups. Moreover, the content which is first shared in these groups soon reaches a wide audience of not previously considered deviant users.

The proliferation of the Internet has transformed child sexual abuse into a crime without geographical boundaries. Child sex offenders turning to the Internet as a means of creating and distributing child pornography has allowed the creation of a network of support groups for child sex offenders, when historically, this was an offence that occurred in isolation \cite{g2016criminal}. This concern was echoed by Mitchell et al. \cite{mitchell2010use}, who recognised that a small percentage of offenders used social networking sites (SNS) to distribute child pornography. While there is scientific debate on whether the online predator is a new type of child sex offender \cite{quayle2000internet} or if those with a predisposition to offend are responding to the opportunities afforded by the new forms of social media \cite{cooper1998sexuality}, empirical evidence points to the problem of Internet-based paedophilia as endemic. Recent work, such as \cite{winters2017stages,zambrano2019does}, shows that nearly half of the offenders who had committed one or more contact offences, i.e., they had directly and physically abused children, had displayed so-called ``grooming behaviour''.

However, when investigating the possibility to develop automated methods to detect grooming online, researchers are confronted with a number of issues. First, there is only one benchmark dataset available that contains (English) chat conversations written by child sex offenders, the PAN 2012 Sexual Predator Identification dataset, which leverages data from PJ. Concretely, PJ data comprises a single class of chats in the context of PAN 2012 data. Yet, because the victims were actually adult volunteers posing as children, it is likely that these conversations are not entirely representative of online predator-victim communications \cite{pendar2007toward}. Moreover, since the seduction stage often shows similar characteristics with adults' or teenagers' flirting, initial studies trying to detect predatory behaviour directly on the user level typically resulted in numerous false positives when they were applied to non-predatory sexually-oriented chat conversations in the PAN 2012 dataset \cite{inches2012overview}.

For machine learning algorithms to be effective in identifying online sexual predators, they need to be trained with both illegal conversations between offenders and their victims and sexually-oriented conversations between consenting adults \cite{pendar2007toward}. Since such data are rarely made public, initial studies \cite{pendar2007toward,mcghee2011learning} only experimented with data from PJ. The k-NN classification experiments based on word token n-grams performed in \cite{pendar2007toward} achieved up to 93.4\% F-score (trigrams with
k = 30) when identifying the predators from the pseudo-victims. Miah et al. \cite{miah2011detection} were the first to include additional corpora in the non-predator class. They included 85 conversations containing adult descriptions of sexual fantasies and 107 general non-offensive chat logs from websites like \url{http://www.fugly.com} and \url{http://chatdump.com}. When distinguishing between 200 PJ conversations and these additional chat logs, the Na\"ive Bayes classifier outperformed the Decision Tree and the Regression classifier, which resulted in an F-score of 91.7\% for the PJ class. In \cite{bogdanova2014exploring,peersman2012conversation,morris2012identifying,hidalgo2012combining}, the researchers used a corpus of cybersex chat logs and the Naval Postgraduate School (NPS) chat corpus and experimented with new feature types such as emotional markers, emoticons and imperative sentences and computed sex-related lexical chains to detect offenders directly in the PJ dataset automatically. Their Naïve Bayes classifier yielded an accuracy of 92\% for PJ predators \textit{vs} NPS and 94\% for PJ predators \textit{vs} cybersex based on their high-level features. However, both \cite{miah2011detection} and \cite{bogdanova2014exploring} did not filter out any cues that were typical of the social media platforms from which the additional corpora were extracted, which could entail that their models were (to some degree) trained on detecting these cues rather than the grooming content. Moreover, because the high-level features described by Bogdanova et al. \cite{bogdanova2014exploring} were (partially) derived from the PJ dataset itself, these experiments may have resulted in overestimated accuracy when detecting predators from the same dataset.

Recently, the detection of Internet child sex offenders has been extensively investigated in the framework of the PAN 2012 competition, during which efforts have been made to pair the PJ data with a whole range of non-predatory data, including cybersex conversations between adults \cite{inches2012overview}. Because the PAN 2012 benchmark dataset was heavily skewed towards the non-predatory class, most participants applied a two-stage classification framework in which they combined information on the conversation level to the user level \cite{Villatoroe2012:pan}. Moreover, apart from one submission that used character-gram features, all other studies used (combinations of) lexical (e.g., token unigrams) and ``behavioural'' features (e.g., the frequency of turn-taking or the number of questions asked). The best results were achieved by Morris and Hirst \cite{morris2012identifying}, who used a Neural Network classifier combined with a binary weighting scheme in a two-stage approach to first identify the suspicious conversations and, secondly, to distinguish between the predator and the victim. Their system achieved an F-score of 87.3\%. However, during their study, they assumed that ``\textit{predators usually apply the same course of conduct pattern when they are approaching a child}'' \cite{morris2012identifying}, which is in contrast with research by \cite{gottschalk20analysis11dark}, which resulted in three different types of predators and, hence, of grooming approaches. Moreover, the PAN2012 dataset was also not cleansed of platform-specific cues, which could again have led to overestimated F-scores during the competition. A more detailed overview of the results of the PAN 2012 International Sexual Predator Identification Competition can be found in \cite{inches2012overview}.

With regard to the content of predatory chat conversations, McGhee et al. \cite{mcghee2011learning} were the first to investigate the possibility to detect different stages in the grooming process automatically. Based on an expanded dictionary of terms they applied a rule-based approach, which categorised a post as belonging to the stage of gaining personal information, grooming (which included lowering inhibitions or re-framing and sexual references), or none. Their rule-based approach outperformed the machine learning algorithms they tested and reached up to 75.1\% accuracy when categorising posts from the PJ dataset into one of these stages. A similar approach was used by Michalopoulos and Mavridis \cite{michalopoulos2011utilizing}, whose Na\"ive Bayes classifier achieved a 96\% accuracy when categorising predatory PJ posts as belonging to either the gaining access, the deceptive relationship or the sexual affair grooming stage. The second task of the PAN 2012 competition consisted of detecting the specific posts that were most typical of predatory behaviour from the users that were labelled suspicious during the first task. To this end, most participants either created a dictionary-based filter containing suspicious terms \cite{morris2012identifying,paraparlearning} or used their post-level predictions from the predator identification task \cite{kontostathis2012identifying,hidalgo2012combining}. The best F-score was achieved by \cite{peersman2012conversation}, who used a dictionary-based filter highlighting the utterances that referred to one of the following grooming stages: sexual stage, re-framing, approach, requests for data, isolation from adult supervision and age- and child-related references. Their approach resulted in a 35.8\% precision, a 26.1\% recall and a 30.2\% F-score. Finally, Elzinga et al. \cite{elzinga2012analyzing} proposed a method based on Temporal Concept Analysis using Temporal Relational Semantic Systems, conceptual scaling and nested line diagrams to analyse PJ chat conversations. Their transition diagrams of predatory chat conversations seemed to be useful for measuring the level of threat each offender poses to his victim based on the presence of the different grooming stages.

Although these studies showed promising results, the issue remains that these methods are applied to a corpus that contains conversations between offenders and pseudo-victims. Hence, the adult volunteers that were posing as children could not accede to requests for ``cammin'', sending pictures, etc. As a result, the PJ dataset contains hardly any conversations by groomers, because this type of offender typically does not invest much time in the seduction process and switches to a different victim when his needs are not fulfilled quickly. Moreover, it is highly likely that children would have responded differently to the grooming utterances than the adult volunteers did, which could have influenced the language use of the offenders.
\subsection{Latent Dirichlet Allocation}

The Latent Dirichlet Allocation (LDA) is a type of generative probabilistic model proposed by \cite{blei2003latent}. It comprises an endogenous NLP technique, which as highlighted in \cite{cambria2014jumping} ``\textit{involves the use of machine-learning techniques to perform semantic analysis of a corpus by building structures that approximate concepts from a large set of documents}'' without relying on any external knowledge base. LDA, as the name implies, is a latent variable model in which each item in a collection (e.g., each text document in a corpus) is modelled as a finite mixture over an underlying set of topics. Each of these topics is characterised by a distribution over item properties (e.g., words). LDA assumes that these properties are exchangeable (i.e., ordering of words is ignored, as in many other ``bag of words'' approaches in text modelling), and that the properties of each document are observable (e.g., the words in each document are known). The word distribution for each topic and the topic distribution for each document are unobserved; they are learned from the data.

Since LDA is an unsupervised topic modelling method, there is no direct measure to identify the optimal number of topics to include in a model.
What LDA does is to assign to documents probabilities to belong to different topics (an integer number $k$ provided by the user), where these probabilities depend on the occurrence of words which are assumed to co-occur in documents belonging to the same topic (Dirichlet prior assumption). This exemplifies the main idea behind all unsupervised topic models, that language is organised by latent dimensions that actors may not even be aware of  \cite{mcfarland2013differentiating}. Thus, LDA exploits the fact that even if a word belongs to many topics, occurring in them with different probabilities, they co-occur with neighbouring words in each topic with other probabilities that help to define the topics better. The best number of topics is the number of topics that helps the most human interpretability of the topics. This means that if the topics given by LDA can be well-distinguished by humans, then the corresponding number of topics is acceptable. Researchers have recommended various approaches to establish the optimal $k$ (e.g. \cite{cao2009density,arun2010finding,deveaud2014accurate,roder2015exploring,zhao2015heuristic}).
These approaches provide a good range of possible $k$ values that are mathematically plausible. However, according to \cite{dimaggio2013exploiting}, when topic modelling is used to identify themes and assist in interpretation (like in the present study), rather than to predict a knowable state or quantity, there is no statistical test for the optimal number of topics or the quality of a solution. A simple way to evaluate topic models is to look at the qualities of each topic and discern whether they are reasonable \cite{mcfarland2013differentiating}. In addition, the topic number selection was guided by the model's ability to identify a number of substantively meaningful and analytically useful topics. In fact, the increase in fit is sometimes at the expense of interpretability due to overfitting \cite{dyer2017evolution}. Increasing the number of topics, producing ever-finer partitions can result in a less useful model because it becomes almost impossible for humans to differentiate between many of the topics \cite{chang2009reading}. Ultimately, the choice of models must be driven by the questions being analysed. DiMaggio et al. \cite{dimaggio2013exploiting} suggest that the process is empirically disciplined, in that, if the data are inappropriate for answering the analysts' questions, no topic model will produce a useful reduction of the data.
To the best of our knowledge, the topic coherence measure with the largest correlation to human interpretability is the $C_v$ score defined in \cite{roder2015exploring}, which we also adopt in this study to establish the optimal number of topics, see Section 6.

\section{Ethical and legal compliance}

Data scraping from the web is extensively used by academic researchers to track the web,and companies to gain information about their customers. The philosophy of crawling is to index the web and the Internet as a whole, to make information available to the public, and to extract information for different business and research purposes. Yet, due to the invasive practices used for extracting large amounts of information, there is an ongoing debate on the ethical and legal aspects of web data crawling.

According to Internet advocates, if web crawling were to be unethical, then the whole web would not have been discoverable since the entire expansion of the Internet is based on web crawling. As a matter of fact, web scraping has benefited the web so much that virtually everyone on the net is directly or indirectly involved in web scraping. Even big service providers like Google scrap the Internet to be able to provide qualified and verified data in the search results. However, for web data crawling to be ethical, there must be some rules to be followed (like those imposed in the \texttt{robots.txt} file of every web site) to not infringe on the security and the rights of the users. In fact, there are already several professional web scraping service providers who abide by the general rules and regulations to get adequate and appropriate authorisation from the concerned web resource.

As a matter of fact, many scholars advocate that it is the application of the data that have been scrapped and not the web scrapping \textit{per se}, that may be unethical or illegal. For instance, there might be issues when data that are not meant to be made public are scraped and reused for commercial or other purposes. The legal issues of web scraping are widely discussed in the context of the copyrighted and data protection law. The latter is expressed in the EU by the GDPR, which defines the privacy and data protection rights and the rules to be respected when the processing of personal data takes place. While the GDPR is applicable even for research purposes, it states that for meeting ``\textit{the specificities of processing personal data for scientific research purposes, specific conditions should apply in particular as regards the publication or otherwise disclosure of personal data in the context of scientific research purposes}'' (recital 159). Inevitably, when web crawling collects the personal data of web users to facilitate specific research purposes, this processing needs to be aligned with the data protection principles enshrined in the GDPR.

The GDPR requires a specific lawful basis for the processing of the personal data of individuals, with the consent to be the most commonly advertised among them. Beyond consent, however, the GDPR defines some other bases so as the processing of the personal data to be lawful: when the processing is necessary in order to protect the vital interests of the data subject or of another natural person (Article 6(1)(d)); when the processing is necessary for the performance of a task carried out in the public interest (Article 6(1)(e)); or when the processing is necessary for the purposes of the legitimate interests pursued by the controller or by a third party (Article 6(1)(f)). Therefore, while research is not explicitly designated as its own lawful basis for processing, in some cases it may qualify under Articles 6(1)(d)(e)(f) compatible with some of the already foreseen lawful bases. When a controller collects personal data under a lawful basis, Article 6(4) allows it to process the data for a secondary research purpose. Thus, while the GDPR explicitly permits re-purposing collected data for research, it also may permit a controller to collect personal data initially for research purposes, without requiring the data subject's consent.

Furthermore, although research is not mentioned explicitly as a lawful basis for personal data processing, Recital 157 identifies the benefits associated with personal data research, subject to appropriate conditions and safeguards. These benefits include the potential for new knowledge when researchers ``\textit{obtain essential knowledge about the long-term correlation of a number of social conditions}''. The results of the research ``\textit{obtained through registries provide solid, high-quality knowledge which can provide the basis for the formulation and implementation of knowledge-based policy, improve the quality of life for a number of people, and improve the efficiency of social services.}''

Moreover, the GDPR foresees derogations for the secondary processing of personal data for research purposes as long as there is a lawful basis for such processing (Article 5, Recital 50). Article 89 sets out the ``\textit{appropriate safeguards}'' that controllers must implement to further process personal data for research. It mandates controllers explicitly to put in place ``\textit{technical and organisational measures}'' to ensure that they process only the personal data necessary for the research purposes, in accordance with the principle of data minimisation outlined in Article 5(c). Article 89(1) provides that one way for a controller to comply with the mandate for technical and organisational measures is through the deployment of ``\textit{pseudonymisation}.'' Pseudonymisation is ``\textit{the processing of personal data in such a way that the data can no longer be attributed to a specific data subject without the use of additional information, as long as such additional information is kept separately and subject to technical and organisational measures to ensure non-attribution to an identified or identifiable individual}'' (Article 4(3b)).

Taken the above into consideration, one may consider that the case of personal data scraped from social media sites without the consent of the user that added them, may raise serious concerns regarding its ethical and legal consequences. Yet, these concerns can be easily removed, as we demonstrate below, when data scrapping is performed by researchers to facilitate the mitigation of malevolent uses such as those of pedophile and sex exploitations. More specifically, a team of researchers scraped a well-known social media site that attracts millions of teenagers (even if the web site's terms of service forbid its use by people under the age of 18) and they found that the exchanged text chats among its participants include numerous instances of discussions involving sexual harassment and pedophile actions, all covered up under seemingly innocent words and terminologies that are impossible to be tracked by conventional software tailored to identify specific words for sex abuse. To facilitate research on advanced and innovative ways of tracking down suspicious cases of child abuse and harassment, the researchers, after scrapping the chats on the site referring to the coded malevolent conversations, published a dedicated corpus including these suspicious words, strings and emoticons. All user data, namely the user's nickname, have been anonymised with masking techniques whereas every single user was always masked with the same string. Taking into account that the identification of the users could be potentially possible when additional information (held by the researchers) is used, this masking technique is in fact a pseudonymisation in GDPR terms. Since pseudonymised data are still personal, they still fall under the scope of the GDPR. Therefore, researchers had to ensure that the processing of the personal data contained in the scrapped chats is compatible with the data protection provisions of the GDPR, and in particular with at least one of the six lawful purposes of processing enshrined in GDPR Article 6. Taking into account that the undertaken data crawling of the personal data can protect the vital interests of the children participating in the social media site so an not to be fooled by pedophile users, as well as that this processing is beyond any doubt carried out in the public interest, the data scrapping and subsequent analysis of the concerned data by the researcher is in accordance with the GDPR.

Special attention should be paid for the processing of users data, given that the processed information most likely refers to the sexual preferences of the data subjects, a piece of information considered to be among the special categories of personal data referred to as ``\textit{sensitive}'' for which stricter provisions apply (Article 9). Yet, derogating from the prohibition on processing special categories of personal data ``\textit{should also be allowed when provided for in Union or Member State law and subject to suitable safeguards, so as to protect personal data and other fundamental rights, where it is in the public interest to do so}'' (recital 52, Article 9(2)(j)). And beyond any doubt, protecting children from pedophile actions and sexual harassment is, above all, of substantial public interest and has been foreseen to all domestic legislations. Therefore, provided that the processing of the personal data of the data subjects is proportionate to the aim pursued, respects the essence of the right to data protection and provides for suitable and specific measures, i.e. pseudonymisation, to safeguard the fundamental rights and the interests of the data subject, the derogations for processing sensitive information under the Article 9(2) are fulfilled.

Finally, the GDPR Article 12(1) requires controllers to ``\textit{take appropriate measures}'' to inform data subjects of the nature of the processing activities and the rights available to them. Controllers are required to provide this information in all circumstances, regardless of whether consent is the basis for processing, ``\textit{in a concise, transparent, intelligible and easily accessible form, using clear and plain language}'' (Article 12(1)). Nevertheless, a researcher may be exempted from the notice requirement if she received the personal data from someone other than the data subject, such as where the data came from a publicly available source. Article 14 exempts controllers in these circumstances, if ``\textit{the provision of such information proves impossible or would involve a disproportionate effort},'' which ``\textit{could in particular be the case}'' in the research context (Recital 62). A researcher also may claim an exemption if providing notice would be ``\textit{likely to render impossible or seriously impair the achievement of the [research] objectives,'' provided there are appropriate safeguards in place, ``including making the information publicly available}'' (Article 14(5)(b)).

In summary, scrapping personal data from social media sites and publishing them in pseudonymised form for research purposes is legal and ethical as long as it is performed to protect the vital interests of the data subjects or others and it is in the public interest to do so.

\section{The dataset}
\label{sec:dataset}
In what follows, we analyse a large-scale dataset that we created based on the public interactions between streamers and viewers during the live broadcasts of users identified as adult content producers in \cite{lykousas2018adult}, from the LiveMe\footnote{https://www.liveme.com/} platform, a major Social Live Streaming Service (SLSS). The dataset comprises $39,382,838$ chat messages exchanged by $1,428,284$ users, in the context of $291,487$ live broadcasts during a period of approximately two years, from July 2016 to June 2018. Each broadcast effectively functions as a temporary chatroom, where the audience can interact with the streamers via text messages and reward them with virtual rewards, e.g. points, gifts, badges (some of which are purchasable), or even virtual money. Apart from the chat messages, the dataset contains a wide range of user interactions along with metadata. We describe the features below:
\begin{itemize}
\item  \textbf{Metadata (broadcast)}
\begin{itemize}
\item Total Viewers: total number of viewers who joined the livestream as viewers.
\item Duration: duration of stream in seconds
\end{itemize}
\item  \textbf{Metatadata (broadcaster)}
\begin{itemize}
\item Country Code
\end{itemize}
\item  \textbf{Interactions}
\begin{itemize}
\item Likes: Viewers who liked the broadcast \& the number of likes given.
\item Follows: Viewers who followed the broadcaster during the livestream.
\item Gifts: Viewers who sent virtual gifts to the broadcaster, along with value (in virtual currency) for each gift.
\item Shares: Viewers who shared the broadcast (via a link so others can join).
\item Blocks: Viewers who have been blocked by the broadcaster (i.e. banned from a stream).
\end{itemize}
\end{itemize}

\begin{figure*}[!ht]
  \centering
  \begin{subfigure}[t]{0.3\textwidth}
      \centering
    \includegraphics[width=\textwidth]{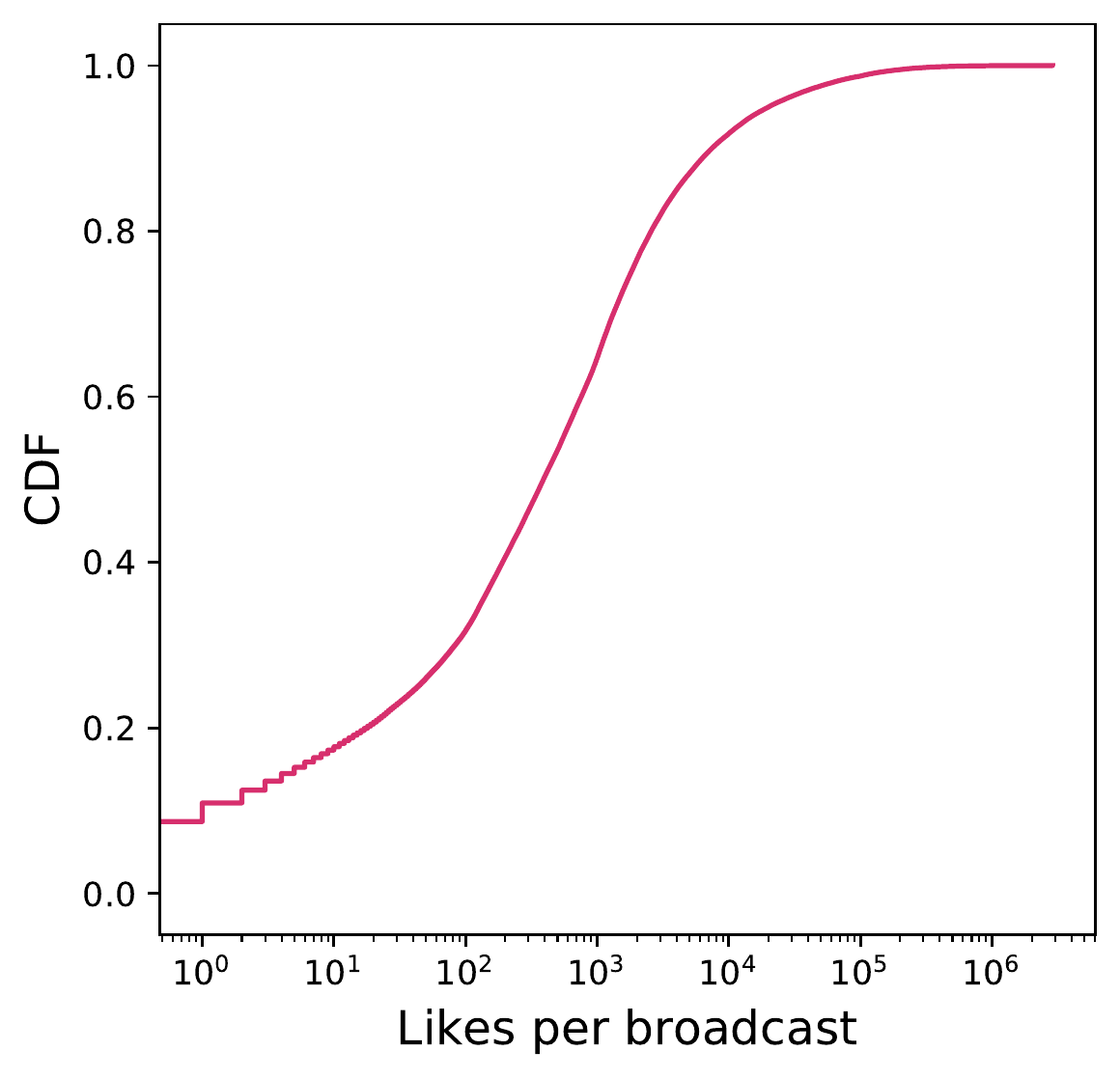}
    \caption{Likes per broadcast}
    \label{fig:likes_cdf}
  \end{subfigure}
  \begin{subfigure}[t]{0.3\textwidth}
      \centering
    \includegraphics[width=\textwidth]{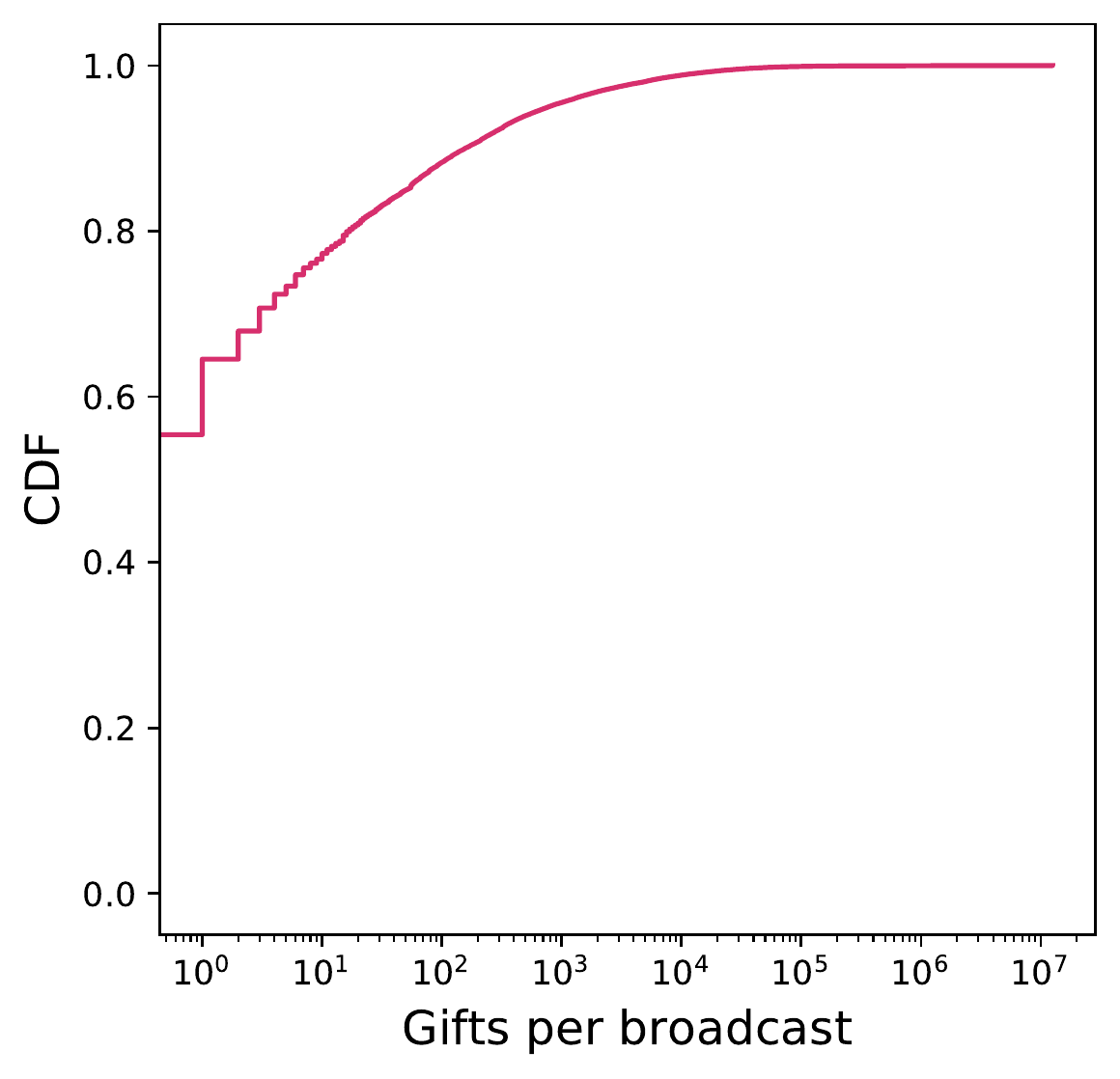}
    \caption{Gifts per broadcast}
    \label{fig:gifts_cdf}
  \end{subfigure}
  \begin{subfigure}[t]{0.3\textwidth}
      \centering
    \includegraphics[width=\textwidth]{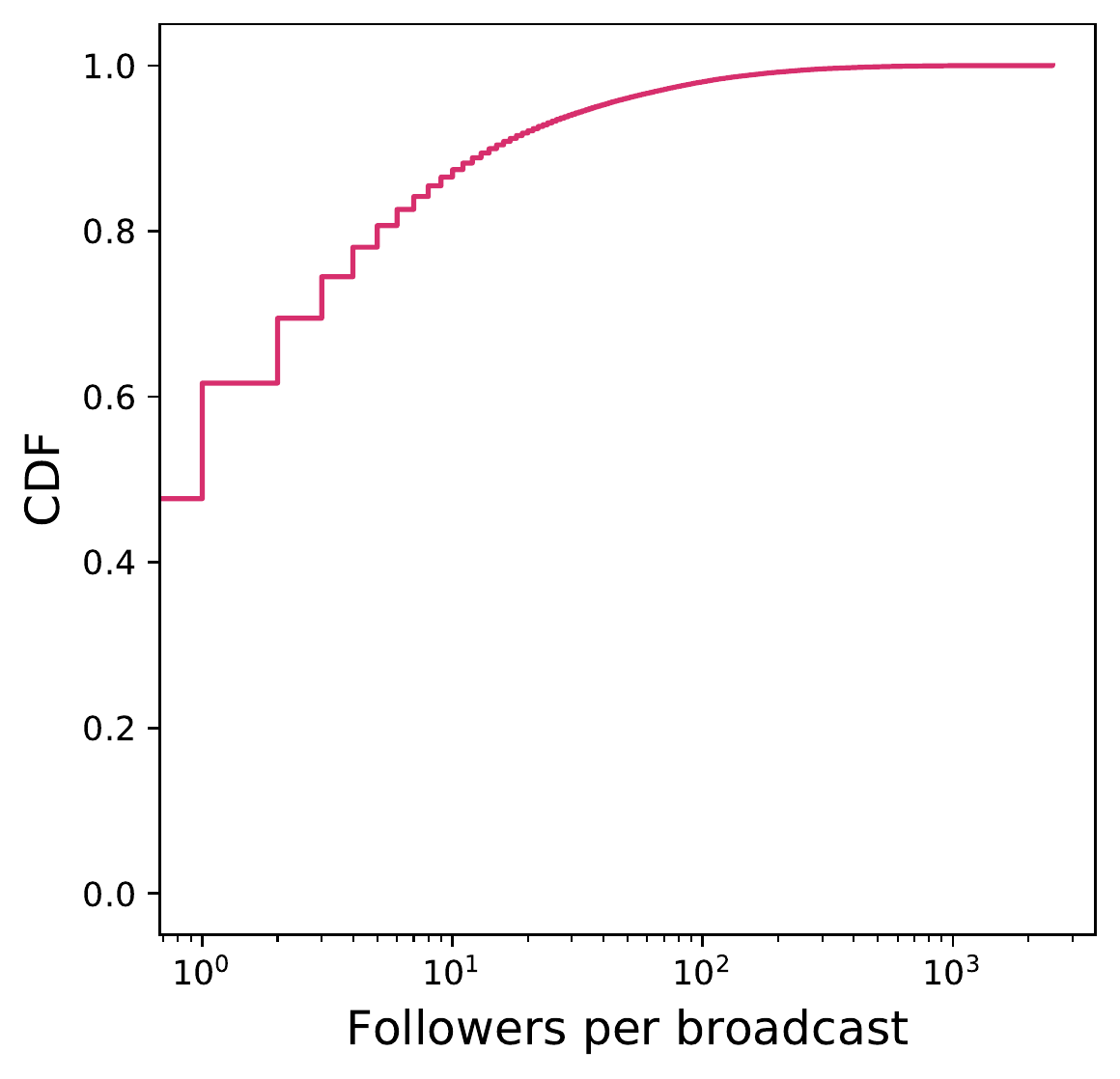}
    \caption{Follows per broadcast}
    \label{fig:follows_cdf}
  \end{subfigure}
  \begin{subfigure}[t]{0.3\textwidth}
      \centering
    \includegraphics[width=\textwidth]{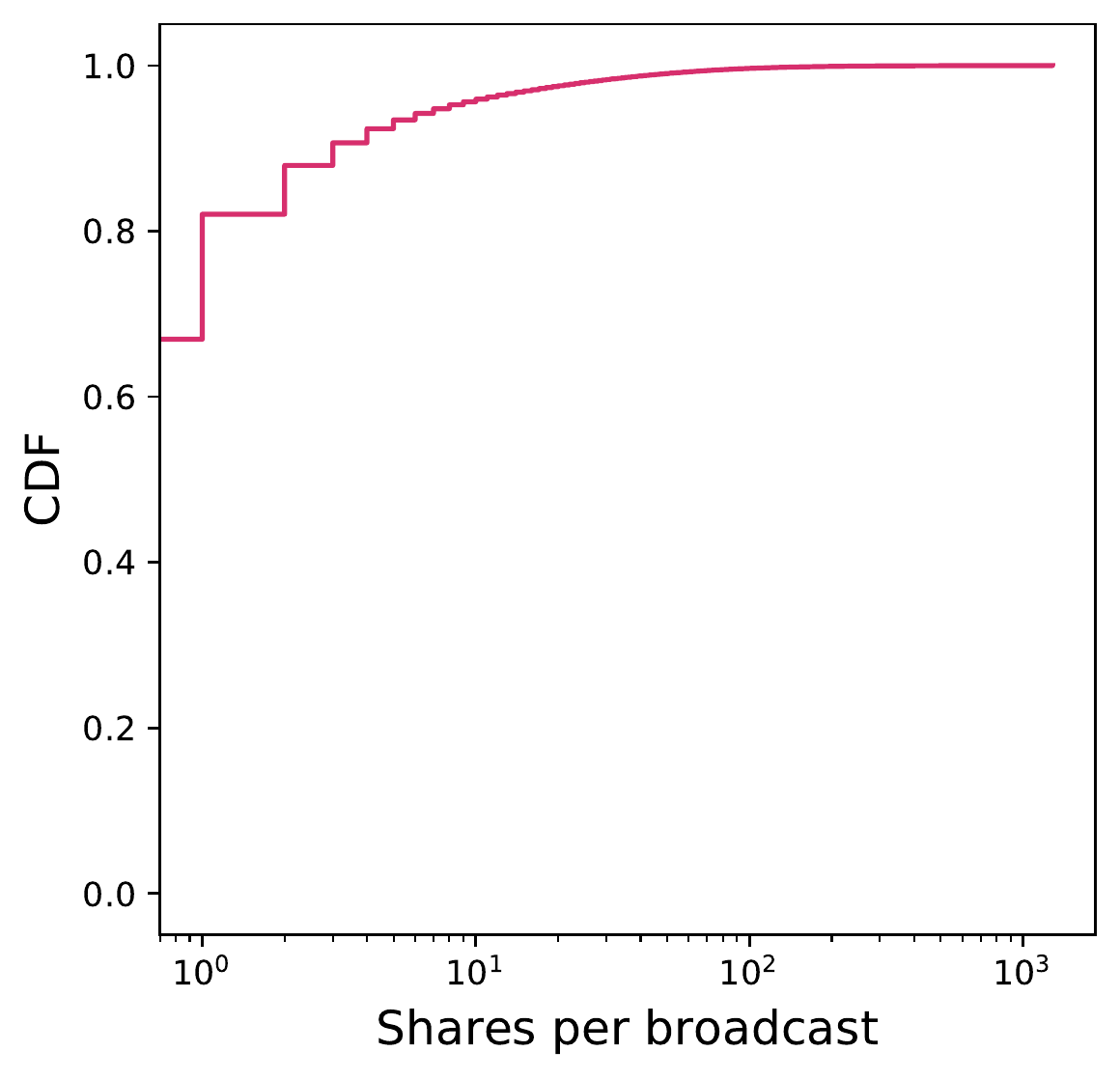}
    \caption{Shares per broadcast}
    \label{fig:shares_cdf}
  \end{subfigure}
  \begin{subfigure}[t]{0.3\textwidth}
      \centering
    \includegraphics[width=\textwidth]{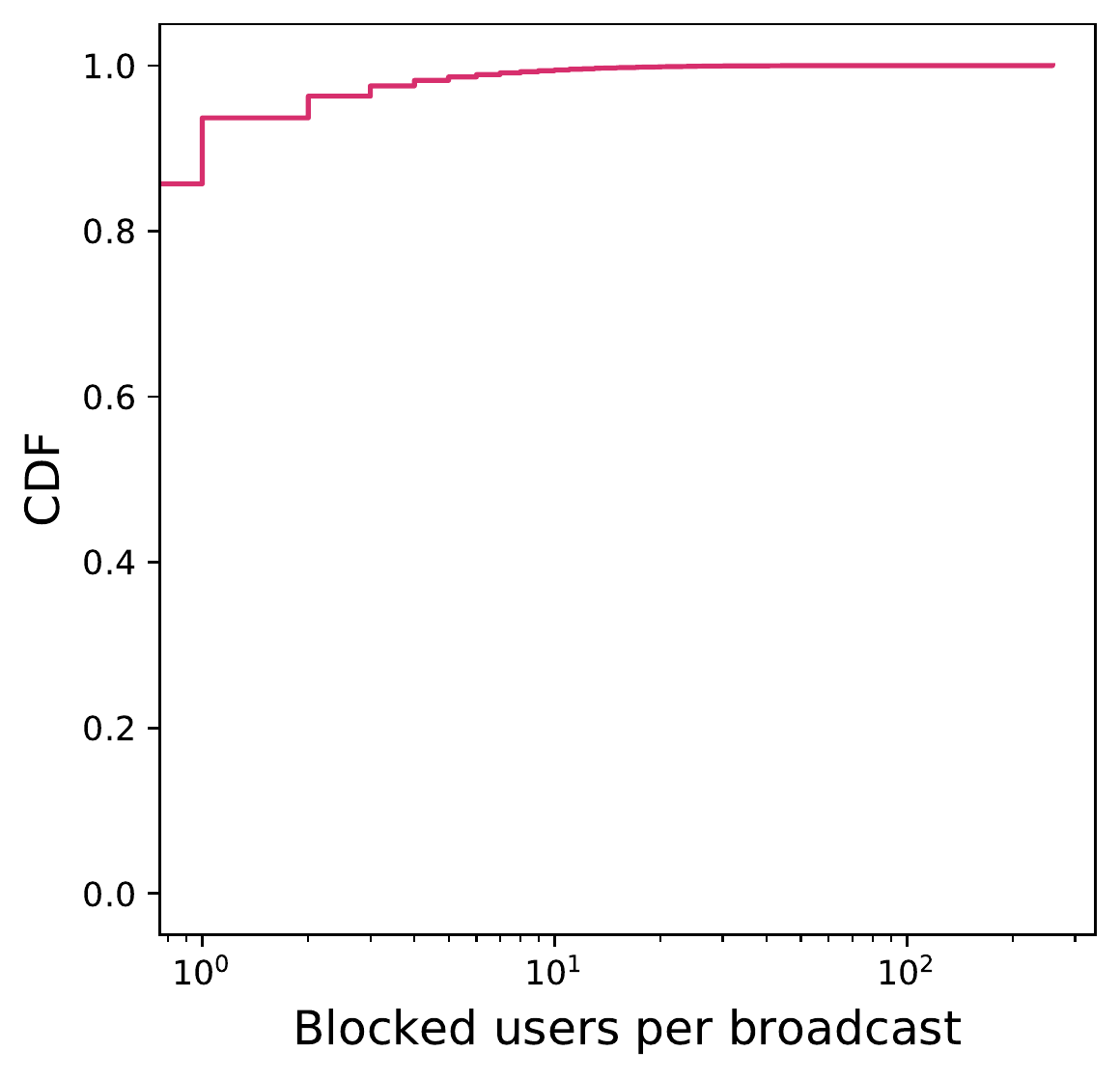}
    \caption{Blocked users per broadcast}
    \label{fig:blocked_cdf}
  \end{subfigure}

  \begin{subfigure}[t]{0.3\textwidth}
      \centering
    \includegraphics[width=\textwidth]{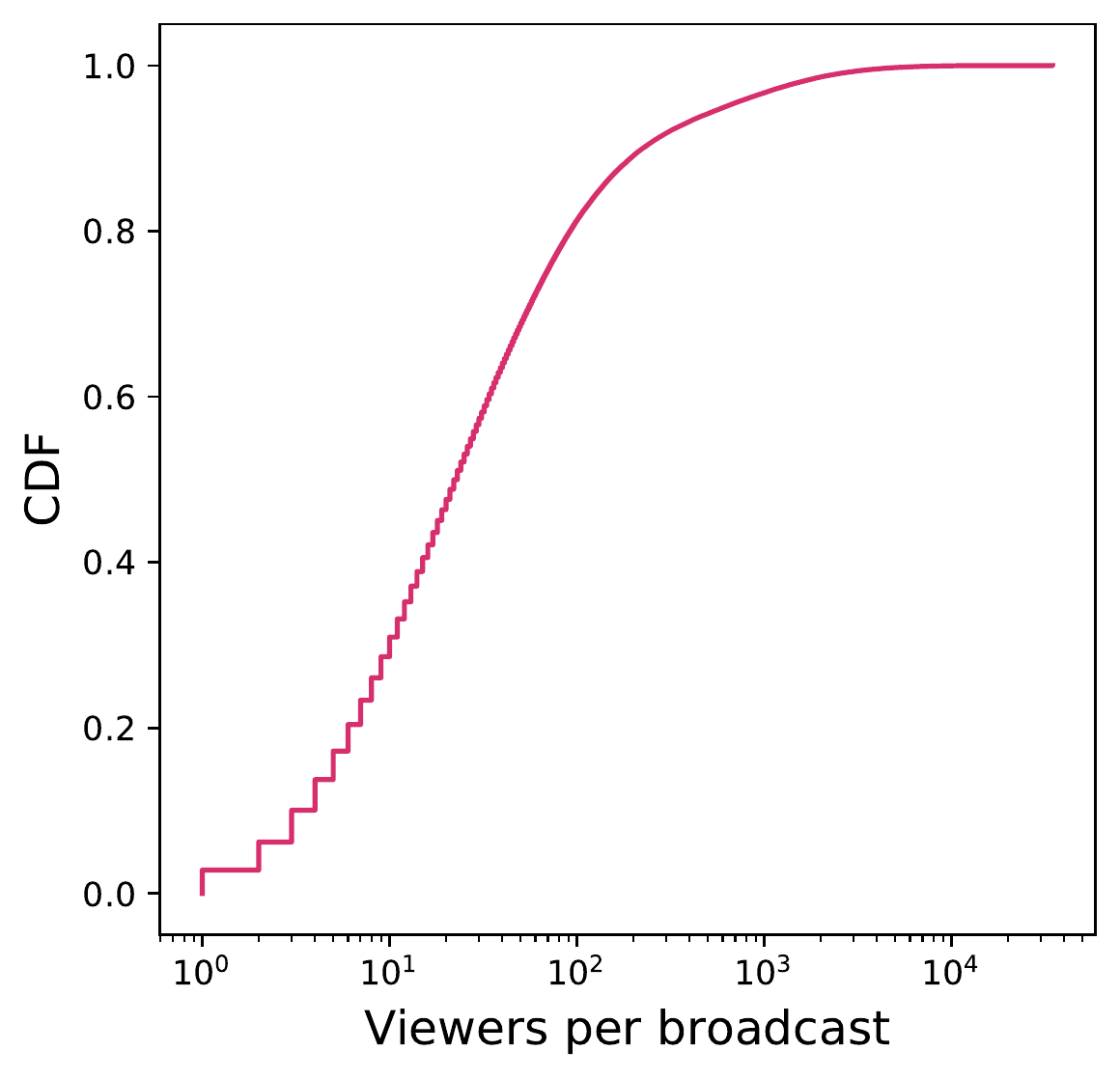}
    \caption{Total viewers per broadcast}
    \label{fig:viewers_cdf}
  \end{subfigure}
  \begin{subfigure}[t]{0.3\textwidth}
      \centering
    \includegraphics[width=\textwidth]{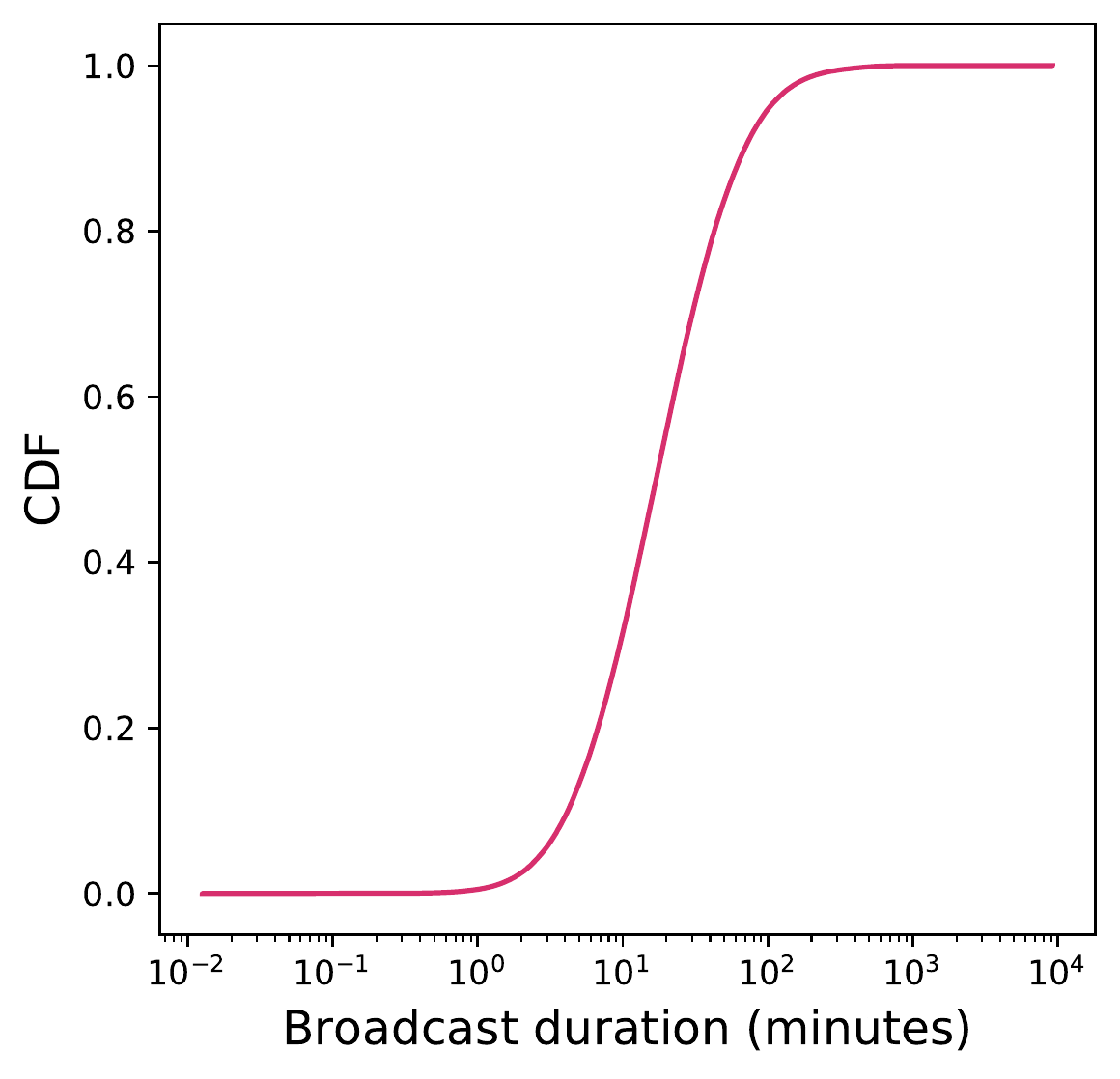}
    \caption{Duration per broadcast}
    \label{fig:duration_cdf}
  \end{subfigure}
  \caption{Cumulative distribution functions (CDFs) of broadcast metadata features and interactions.}
  \label{fig:feature_cdfs}
\end{figure*}

\begin{figure}[!t]
      \centering
   \includegraphics[width=\columnwidth]{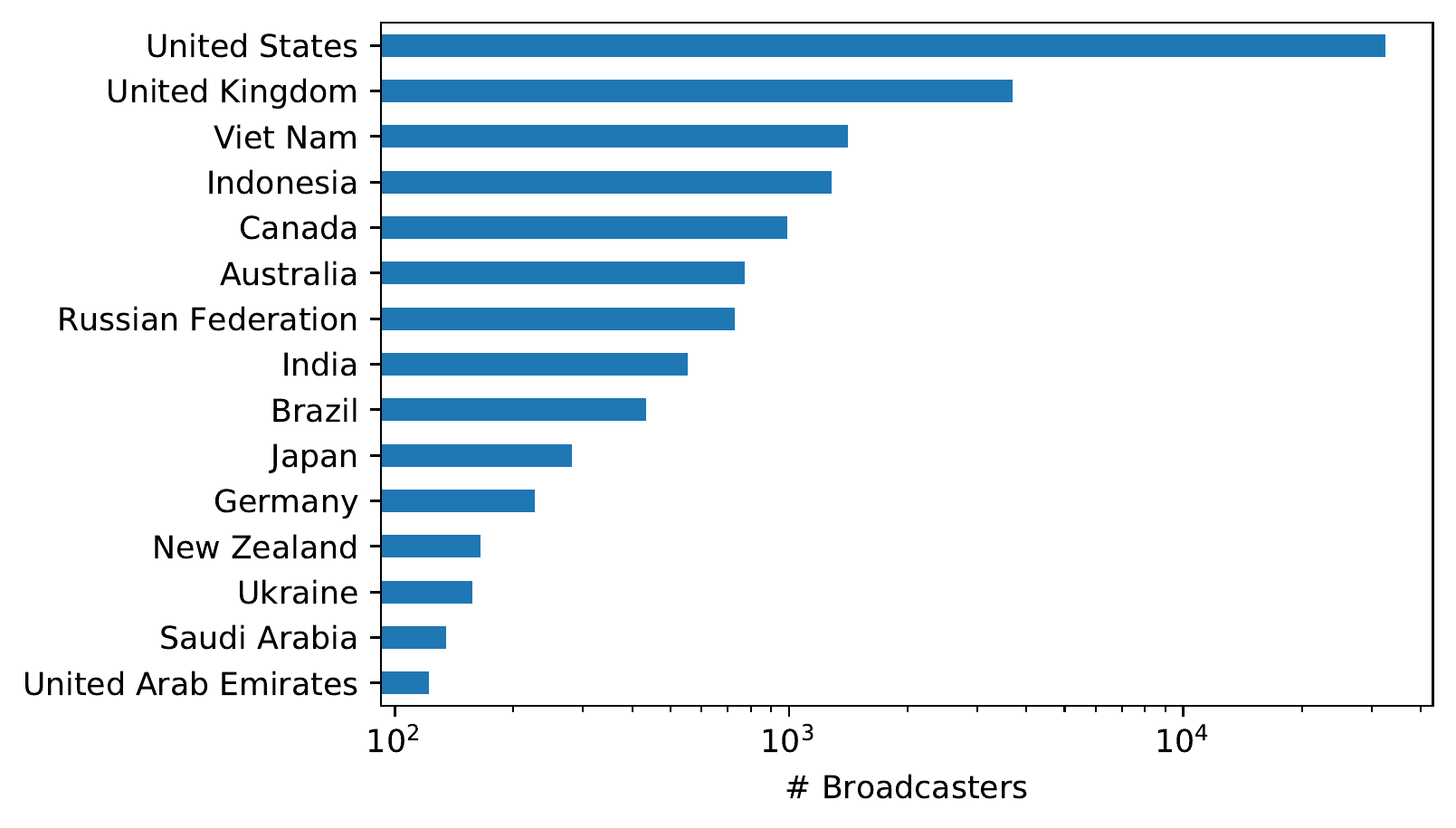}
    \caption{Broadcasters count per country}
    \label{fig:countries}
\end{figure}

In order to get an understanding of how the aforementioned features are distributed, we plot the cumulative distribution functions (CDFs) in Figure \ref{fig:feature_cdfs}. We observe that every broadcast in our dataset had viewers (143.5 in average) and sizeable duration (31.4 minutes in average), and while most of the broadcasts received likes (92\%), the 55\% did not receive any gifts (since they cost money, contrary to likes). Furthermore 47\% for the broadcasts did not generate any new followers for the broadcasters, while the interactions of sharing and blocking are relatively rare in our dataset (i.e. they are zero for 0.67\% and 0.86\% of the broadcasts, respectively).
Next, to get an understanding of the geographical distribution of adult content producers, we plot the distribution of the broadcasters per country of the whole dataset, focusing on the 15 countries with most broadcasters, in Figure \ref{fig:countries}.


\section{Large-scale grooming analysis}
\label{sec:analysis}
\begin{figure}[!ht]
   \begin{subfigure}[t]{0.49\columnwidth}
    \centering
      \includegraphics[width=\columnwidth]{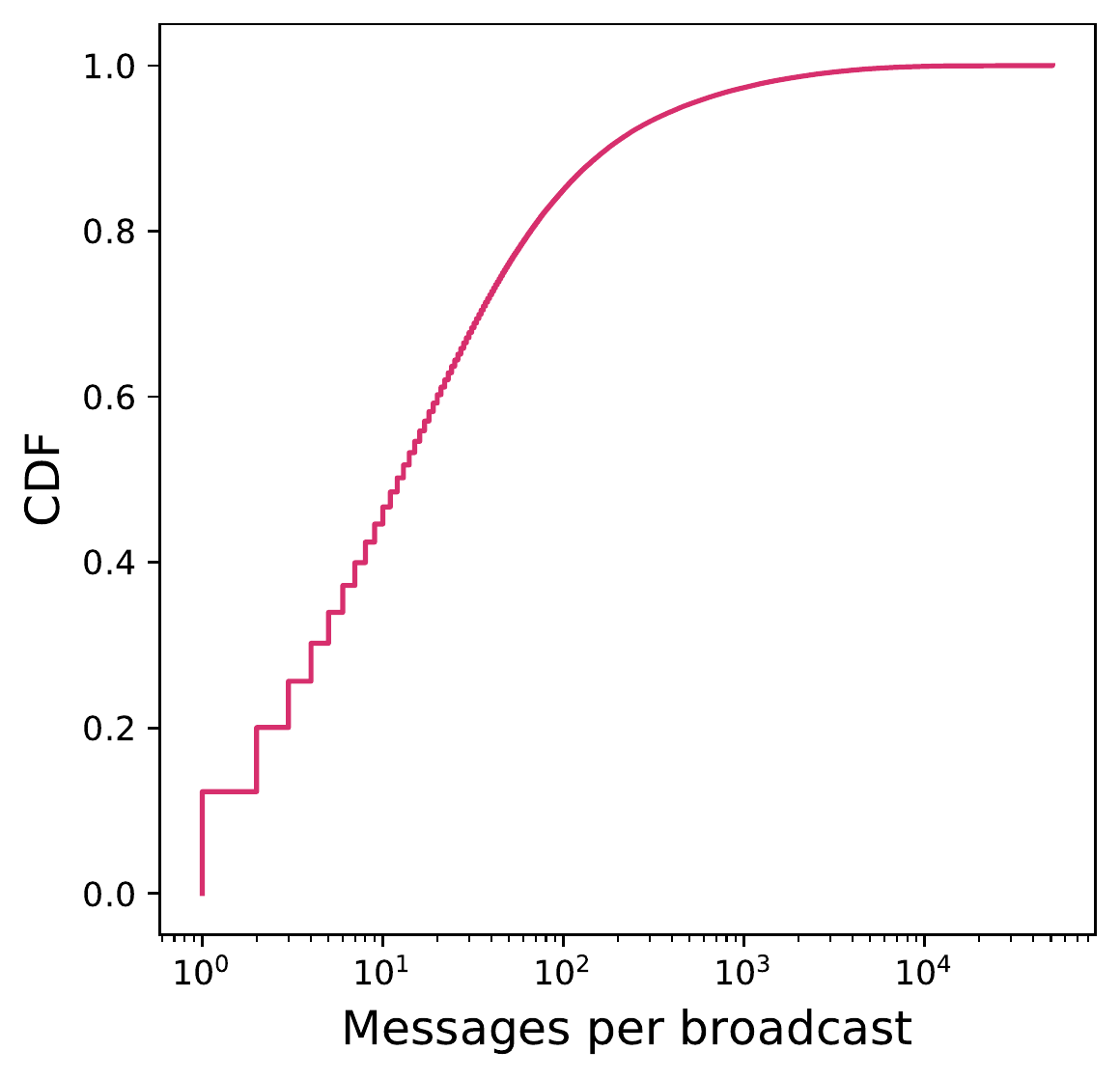}
      \caption{chat messages per broadcast}
      \label{fig:messages_per_chat}
   \end{subfigure}
       \begin{subfigure}[t]{0.49\columnwidth}
    \includegraphics[width=\columnwidth]{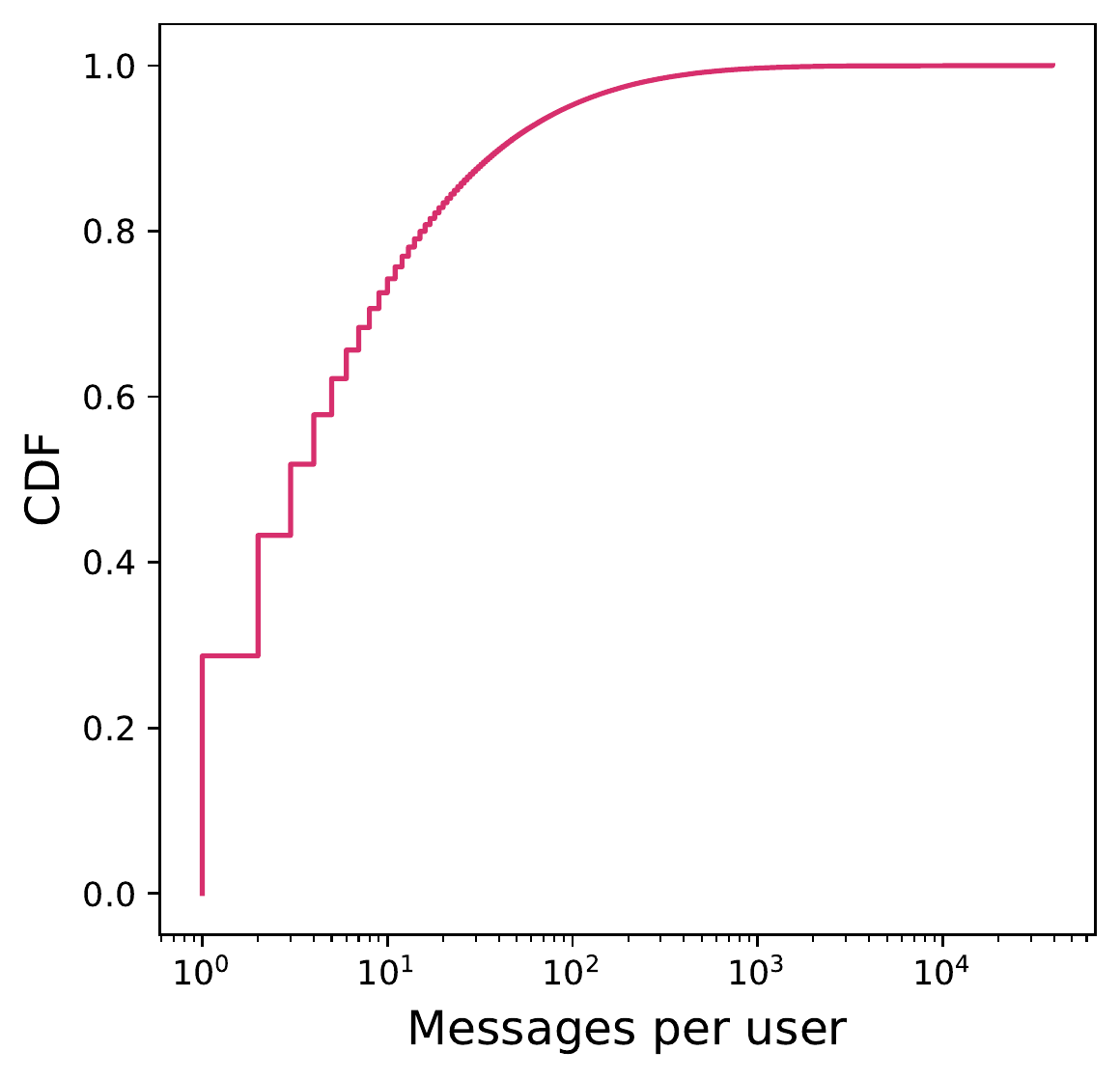}
    \caption{chat messages per user}
    \label{fig:messages_per_user}
    \end{subfigure}
  \caption{Cumulative distribution functions (CDFs) of the chat messages per broadcast and per user.}
  \label{fig:CDFs}
\end{figure}

By plotting the CDFs of the chat messages per broadcast and per user in Figure \ref{fig:CDFs}, we notice that around 82\% of the broadcasts of adult content producers receive less than 100 chat messages. Moreover, out of the unique users chatting during these broadcasts, only 30\% send more than ten messages in total. Both distributions are particularly heavy-tailed, meaning that the majority of chat messages in our dataset are exchanged during a few highly popular broadcasts.

\begin{figure}[!t]
      \centering
   \includegraphics[width=\columnwidth]{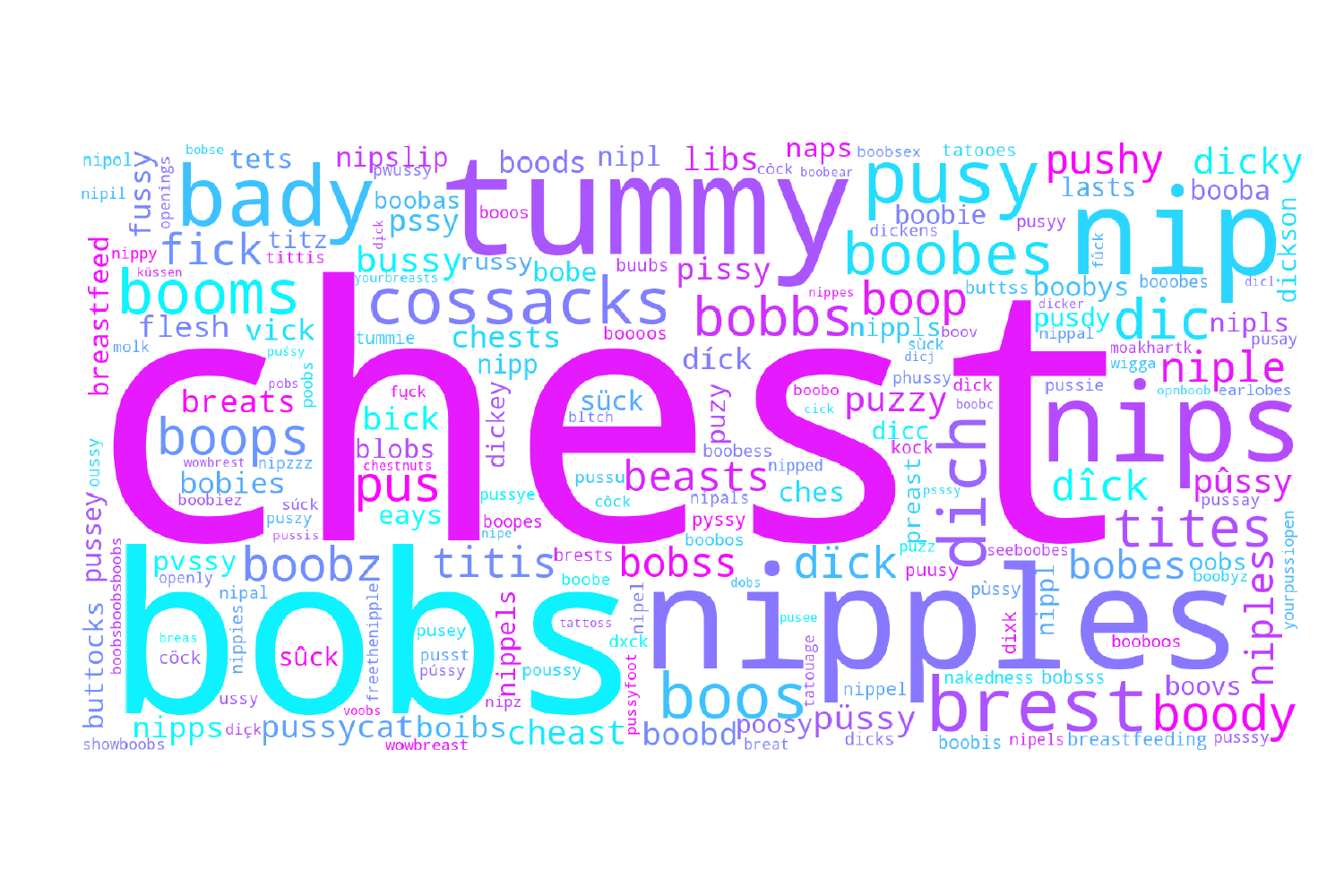}
    \caption{Most frequent semantically-similar words to LIWC sexual terms, as learned by FastText.}
    \label{fig:neighbors}
\end{figure}

To identify sexual grooming behaviour in the chat messages, we adopted the approach followed by several authors in the most recent relevant works \cite{drouin2017linguistic,lorenzo2019so,lorenzo2020communicative} analysing the Perverted-Justice Dataset (PJ), which although dated and relatively small-scale, was the only publicly available dataset of chats produced by online groomers to date.
To this end, we search the chat messages comprising our dataset for sexual content keywords defined in Linguistic Inquiry and Word Count (LIWC) corpus \cite{pennebaker2015development}. More precisely, the 2015 version of the LIWC dictionary for the sexual content variable comprises a total of 131 words. These include a wide range of terms about sexual matters, including sexual orientation (e.g. bi-sexual, heterosexual), sexual organs (e.g. penis*, vagin*, womb), slang terms, sexually transmitted diseases and infections, sexual violence and assault terms and sex enhancements. The most frequently occurring sexual terms in the PJ dataset, had a very low number of occurrences in the LiveMe chats (less than five exact matches in most occasions). The very low occurrence of such words implies the existence of an automated filtering mechanism in place. Nonetheless, relevant literature about online chat has demonstrated that users with previous exposure to text-based automatic moderation techniques can easily circumvent them by introducing noise such as typos, grammatical errors, uncommon abbreviations and out-of-vocabulary words \cite{papegnies2019conversational,hosseini2017deceiving}. To determine whether this is relevant in our dataset, we use Facebook's FastText library \cite{bojanowski2017enriching} to train subword-informed word representations on the LiveMe chats, which we then leverage to identify the semantically-similar adversarial misspellings of filtered terms (such as pussy, boobs, dick, etc.), by querying their nearest neighbours. Our results indicate that indeed this is the case in LiveMe chats, as illustrated in Tables \ref{tab:psy} and \ref{tab:bob}. To illustrate the sexual word misspellings better, we plot the word cloud of the closest neighbours for the relevant LIWC terms in Figure \ref{fig:neighbors}.

\begin{figure}[ht]
      \centering
   \includegraphics[width=\columnwidth]{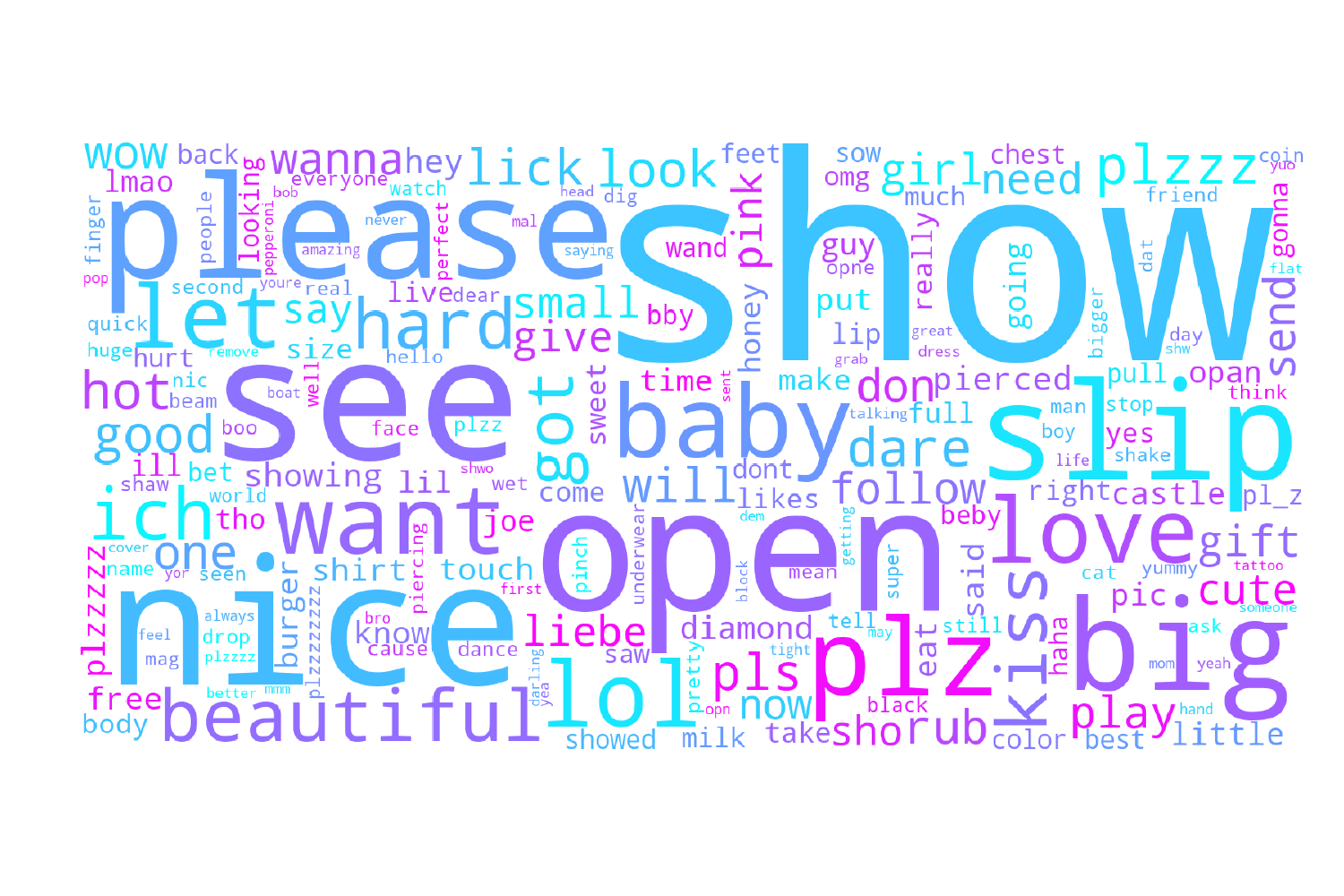}
    \caption{Top collocates of sexual words in LiveMe dataset.}
    \label{fig:collocates}
\end{figure}

\begin{figure}[ht]
      \centering
   \includegraphics[width=\columnwidth]{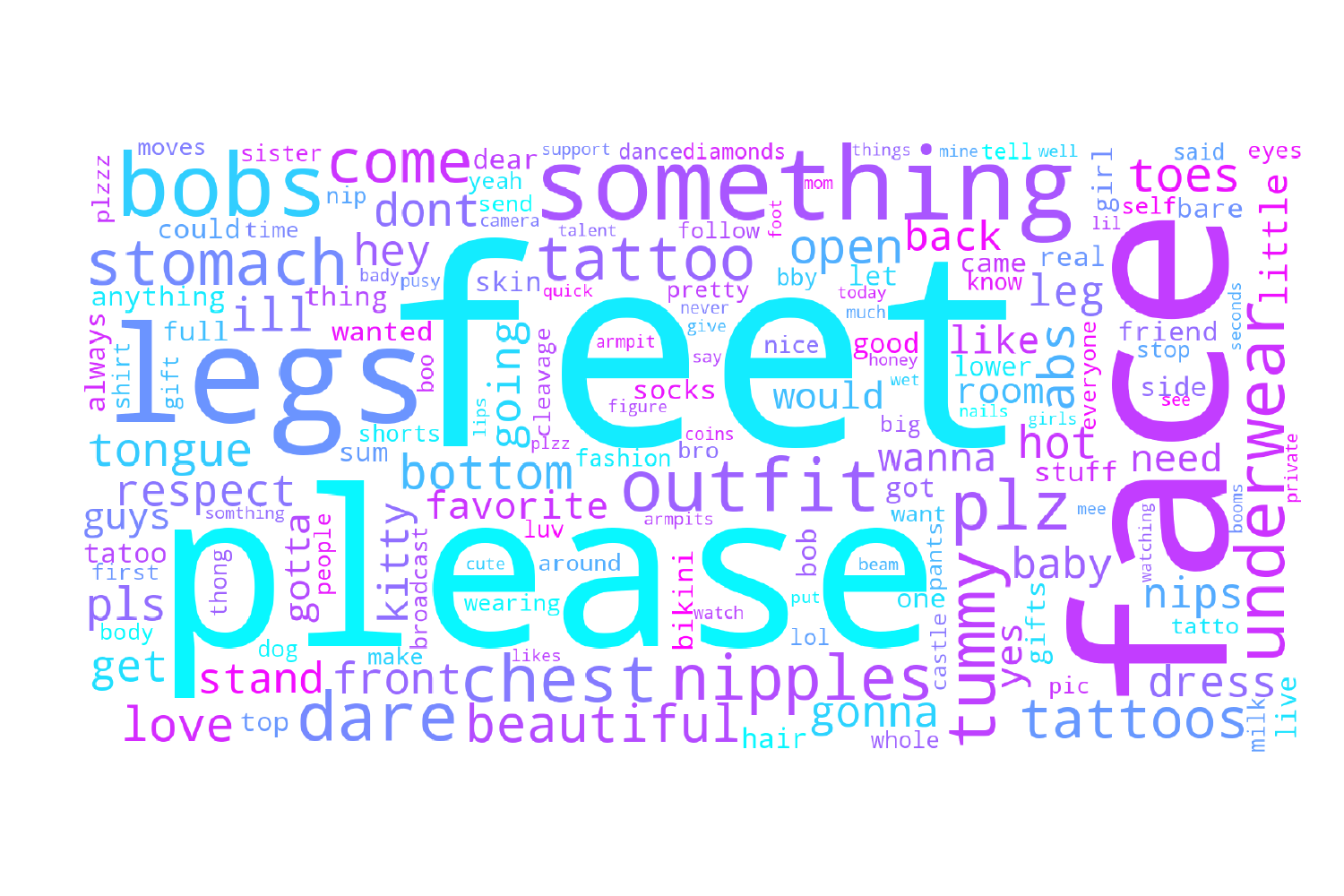}
    \caption{Collocates of the word ``show''.}
    \label{fig:show_collocates}
\end{figure}

Next, to understand the contexts where the aforementioned terms are used, we plot the word cloud of their top collocates in Figure  \ref{fig:collocates}. We notice that the 3 most frequently collocated words are the verbs \textit{show} ($13,329$ collocation occurrences), \textit{open} ($3,032$ collocation occurrences), and \textit{see} ($3,028$ collocation occurrences). To further investigate the imperative meaning of such words in the context of the grooming problem, in Figure \ref{fig:show_collocates}, we plot the top collocates in the whole dataset of chat messages for the most frequent one: \texttt{show} ($203,230$ total occurrences), clearly indicating the existence of sexually predatory behaviours. Similarly, the word \texttt{open} is most frequently collocated with words denoting positive politeness (such as please, plz) and endearment (e.g. baby, dear), as well as sexually connoted words, mostly related to clothing (e.g. underwear, clothes, top, shirt, pants, dress), and emojis representing clothing items (e.g. \includegraphics[height=1em]{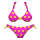},\includegraphics[height=1em]{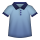},\includegraphics[height=1em]{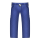},\includegraphics[height=1em]{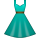}).

\begin{figure}[ht]
      \centering
   \includegraphics[width=\columnwidth]{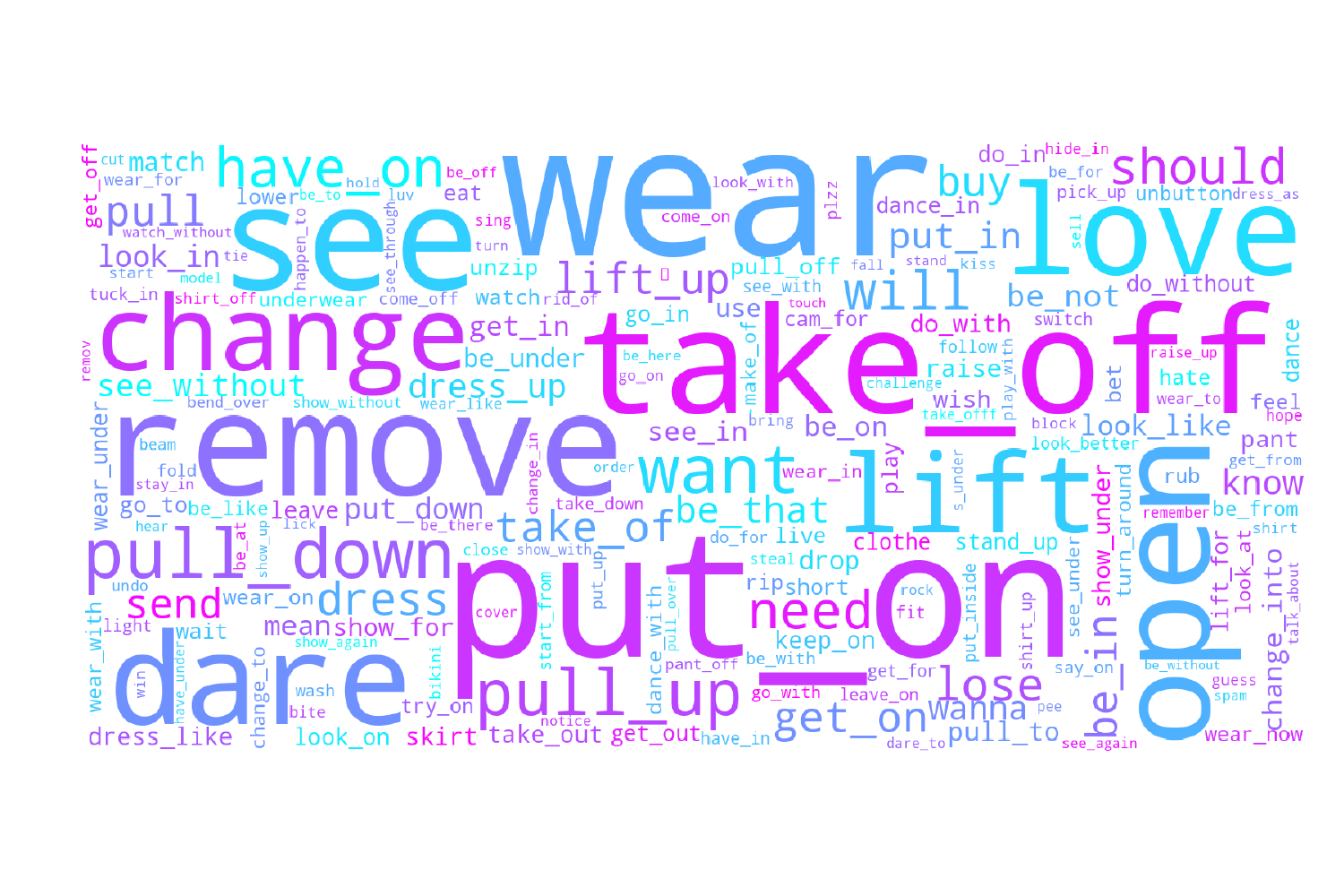}
    \caption{Verbs extracted from chats containing clothing terms.}
    \label{fig:clothing_imperatives}
\end{figure}

While emojis are present in many published datasets, to the best of our knowledge, this is the first study to highlight their relevance in the context of grooming, especially the ones referring to clothing. To this end, we use the same embeddings-based approach as previously described to capture similar clothing terms, see Table \ref{tab:clothes}. Using this method, we assemble a list of 300 unique terms, appearing in the chat messages of 45,086 live streams. Next, to examine the intentions underlying these messages, we performed dependency parsing on every chat message the clothing terms appear in, using the spaCy parser \cite{honnibal2015improved}. From the extracted parse trees, we collected the simple and phrasal verbs. Table \ref{tab:clothing_verbs} contains the 15 most frequently occurring simple and phrasal verbs in their base forms obtained using spaCy lemmatizer\footnote{\url{https://spacy.io/api/lemmatizer}}, after removing the verbs contained in the NLTK \cite{loper2002nltk} stopwords list (e.g. be, can, do, have) to reduce noise in the results.
We plot a word cloud of the extracted verbs and verb phrases in Figure \ref{fig:clothing_imperatives}. The latter, is a clear indication that predators are requesting streamers to perform inappropriate acts involving the removal of their clothes. 
These findings highlight the imperative nature of the predators' communications related to clothing items.

\begin{figure}[!t]
      \centering
   \includegraphics[width=.8\columnwidth]{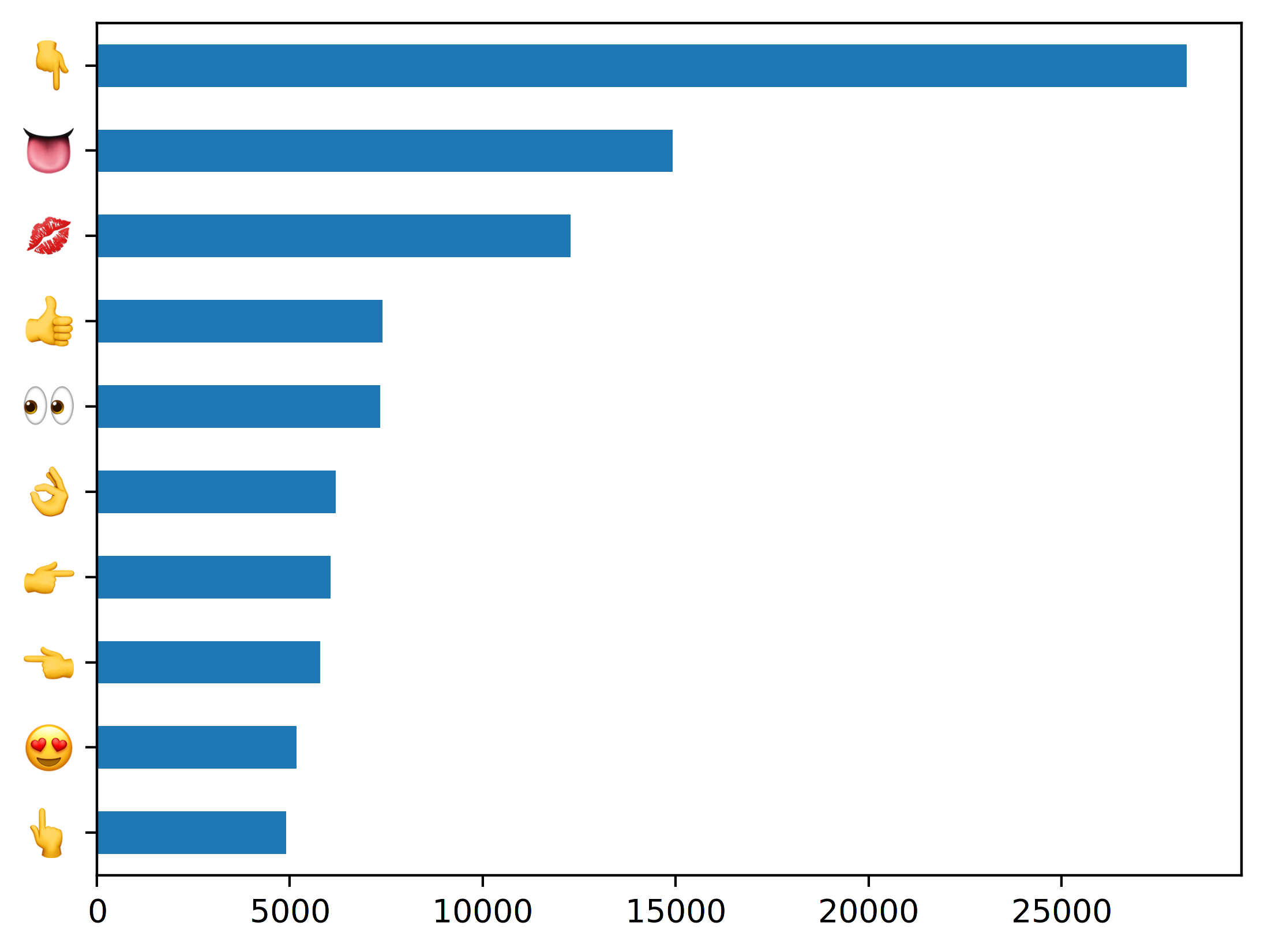}
    \caption{Top emoji collocations for clothing related emojis.}
    \label{fig:emojis}
\end{figure}

Additionally, we explore the use of clothing-related emojis\footnote{\url{https://unicode.org/emoji/charts-12.0/emoji-ordering.html\#clothing}} in our dataset, occurring in 153,797 chats. We note that 83.6\% (128,604) of these messages contain only emojis, without any text. We extract the singular emojis co-occurring with clothing-related emojis since it has been shown that in text messages, emoji sequences tend to have a high level of repetition \cite{mcculloch2018emoji}. Plotting the 10 most frequent emojis, see Figure \ref{fig:emojis}, we may observe that first came the ``backhand index pointing down'' (\protect\includegraphics[height=1em]{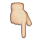}) emoji with 28,248 occurrences, followed by the ``tongue'' emoji (\protect\includegraphics[height=1em]{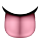}) appearing 14,919 times. Considering the high co-occurrence of emojis depicting hand gestures, we speculate that the use of such emoji combinations comprises a novel nonverbal communication pattern adopted by predators to convey to potential victims their requests for sexually inappropriate and suggestive acts, involving the removal of clothes.

\begin{table}[!t]
    \centering
    \rowcolors{2}{gray!25}{white}
    \begin{tabular}{lrlrlr}
\toprule
   \textbf{Verb} &  \textbf{Count} \\
\midrule
   wear &   6553 \\
   show &   5817 \\
 remove &   3947 \\
    see &   3765 \\
    get &   3157 \\
   open &   3105 \\
   like &   2914 \\
   love &   2913 \\
   dare &   2844 \\
   lift &   2159 \\
 change &   2154 \\
   want &   1811 \\
   take &   1412 \\
     go &   1267 \\
    say &   1237 \\
\bottomrule
\end{tabular}
\rowcolors{2}{gray!25}{white}
\begin{tabular}{lr}
\toprule
   \textbf{Verb} &  \textbf{Count} \\
\midrule
      put\_on &  6083 \\
    take\_off &  4229 \\
   pull\_down &  1930 \\
     pull\_up &  1625 \\
     have\_on &  1214 \\
     take\_of &   731 \\
      get\_on &   728 \\
     lift\_up &   690 \\
      put\_in &   507 \\
    dress\_up &   433 \\
 see\_without &   371 \\
     look\_in &   336 \\
    put\_down &   312 \\
   look\_like &   295 \\
 change\_into &   274 \\
\bottomrule
\end{tabular}
\caption{Top 15 verbs (simple and phrasal) associated with clothing items.}
\label{tab:clothing_verbs}
\end{table}

\begin{table}[!t]
    \centering
    \rowcolors{2}{gray!25}{white}
\begin{tabular}{lcccc}
\toprule
     \textbf{Term} &  \textbf{Distance} &  \textbf{Count} &  \textbf{\#Broadcasts} &  \textbf{\#Users} \\
\midrule
     pusy &  0.828122 &    956 &         513 &     432 \\
      pus &  0.768473 &    416 &         305 &     259 \\
    pushy &  0.741119 &    267 &         185 &     158 \\
    bussy &  0.799563 &    209 &         128 &     100 \\
    püssy &  0.810713 &    198 &         133 &     101 \\
    puzzy &  0.753680 &    195 &         122 &     113 \\
    pûssy &  0.781377 &    184 &         110 &      90 \\
 pussycat &  0.818996 &    169 &         138 &     141 \\
    pissy &  0.702024 &    160 &         141 &     142 \\
     pssy &  0.812888 &    135 &         103 &      79 \\
\bottomrule
\end{tabular}
\caption{Top 10 nearest neighbors (cosine distance) of the word ``pussy''.}
\label{tab:psy}
\end{table}

\begin{table}[!t]
    \centering
\begin{tabular}{lrrrr}
\toprule
   \textbf{Term} &  \textbf{Distance} &  \textbf{Count} &  \textbf{\#Chatrooms} &  \textbf{\#Users} \\
\midrule
   bobs &  0.752709 &  14728 &        5720 &    5754 \\
   boos &  0.756812 &    670 &         490 &     444 \\
  booms &  0.759904 &    638 &         305 &     189 \\
 boobes &  0.868892 &    578 &         315 &     182 \\
  bobbs &  0.794095 &    494 &         341 &     292 \\
  boops &  0.803665 &    452 &         276 &     177 \\
  boody &  0.784702 &    400 &         285 &     190 \\
  boobz &  0.858590 &    389 &         256 &     161 \\
  bobss &  0.787802 &    267 &         175 &     113 \\
  boobd &  0.896997 &    159 &         146 &     150 \\
\bottomrule
\end{tabular}
    \caption{Top 10 nearest neighbors (cosine distance) of the word ``boobs''.}
\label{tab:bob}
\end{table}

\begin{table}[!t]
    \centering
    \rowcolors{2}{gray!25}{white}

\begin{tabular}{lrrr}
\toprule
      \textbf{Term} &  \textbf{Count} &  \textbf{\#Chatrooms} &  \textbf{\#Users} \\
\midrule
     shirt &  36306 &   19070 &   16969 \\
    shorts &  17449 &    7635 &    7635 \\
     dress &  12319 &    7267 &    6154 \\
     pants &  11693 &    6597 &    6479 \\
     short &  10682 &    6379 &    6416 \\
   clothes &  10504 &    5940 &    5905 \\
 underwear &   6055 &    2490 &    3022 \\
    bottoms &   4768 &    2997 &    3272 \\
    bikini &   4621 &    1928 &    1993 \\
     socks &   4563 &    1855 &    2581 \\
\bottomrule
\end{tabular}
\caption{Top 10 most frequent clothing terms.}
\label{tab:clothes}
\end{table}

\section{Topic modelling}
\label{sec:lda}
In this section, we investigate the extent to which grooming behaviours can be modelled mainly using the textual content of chat messages in broadcasts. To this end, we consider a class of probabilistic techniques called ``topic models'', comprising a method well suited to the study of high-level relationships between text documents.



In this study, all the chat messages sent by users during a broadcast are considered to represent a \textit{document}, similar to the notion of chat log documents; described in \cite{basher2014analyzing}.
The topics learned from LDA trained on the chat log documents from our dataset could highlight specific terms associated with latent
communication patterns emerging within the broadcasts, that will help us to understand and identify the modus operandi of sexual groomers in the context of SLSS better, by providing meaningful interpretations of different aspects of user behavior within the chats. Additionally, we investigate the connection of user interactions beyond chatting to grooming, to shed light on the mechanics of sexual predatory behaviors in SLSS.

\subsection{Preprocessing}

To reduce noise and variation in the text data, we focus only on the chat messages produced by streamers in English-speaking countries appearing in our dataset, i.e. United States, Great Britain, Australia, Canada and New Zealand. This resulted in 209,624 broadcasts, produced by 38,099 unique users. Next, for each of the selected broadcasts, we preprocess the content of each chat message individually, according to the following procedure: First, we apply standard text-normalisation techniques, including tokenization, whitespace trimming, capital-letter reduction, and discarding tokens of lengths $>15$ and $<2$. Next, provided the prevalence of misspellings related to sexual or clothing terms, as well the abundant use of emojis, we collect the 100 most semantically similar neighbours of each LIWC sexual term, by querying the learned FastText model (417 terms in total), with a single token \textit{SEX\_TERM}. We repeat the same procedure for the clothing-related terms (see Table \ref{tab:clothes}), collecting 334 terms in total, to which we add the clothing-related emojis, as previously described since they are relevant for our analysis. Any term occurring in the set of clothing terms and emojis is similarly replaced by a single token, namely \textit{CLOTHING\_TERM}. To further reduce the noise of the chat data, we repeat the same by substituting the 100 most semantically similar neighbours of the words \texttt{show} and \texttt{open}, which as previously discussed comprise the top collocates for sexual terms. Furthermore, we remove English stopwords defined in the NLTK \cite{loper2002nltk} stopwords list, and we additionally detect and remove \textit{gibberish} text, i.e. character sequences that do not reflect a real word, but they are like a random compilation of characters instead. This is a common spamming behaviour, e.g. misbehaving users clogging online communication channels with gibberish \cite{yin2009detection}. More precisely, the detection of gibberish strings is handled by a software library by Rob Neuhaus\footnote{\url{https://github.com/rrenaud/Gibberish-Detector}}, implementing a two-state Markov chain which learns how likely two characters of the English alphabet are to appear next to each other. The training of the model is done on a large-scale corpus consisting of English texts available on the Project Gutenberg\footnote{\url{https://www.gutenberg.org/}}. A previous study \cite{doll2019automated} assessed the gibberish detection performance of the library and reported an F1-Score of 0.90, which we consider sufficient for the scope of this work.
For the remaining words found in each chat message, we obtain their base forms using the spaCy lemmatizer.
Finally, we discard broadcasts with less than ten messages, since short texts usually contain few meaningful terms, and thus the word co-occurrence information is difficult to be captured by conventional topic models like LDA \cite{hong2010empirical,zhao2011comparing}.
After following these steps, our dataset was reduced to a total of 64,104 broadcasts (30\% of all streams produced by English-speaking broadcasters).

\subsection{LDA models}
\begin{figure}
    \centering
    \includegraphics[width=.7\textwidth]{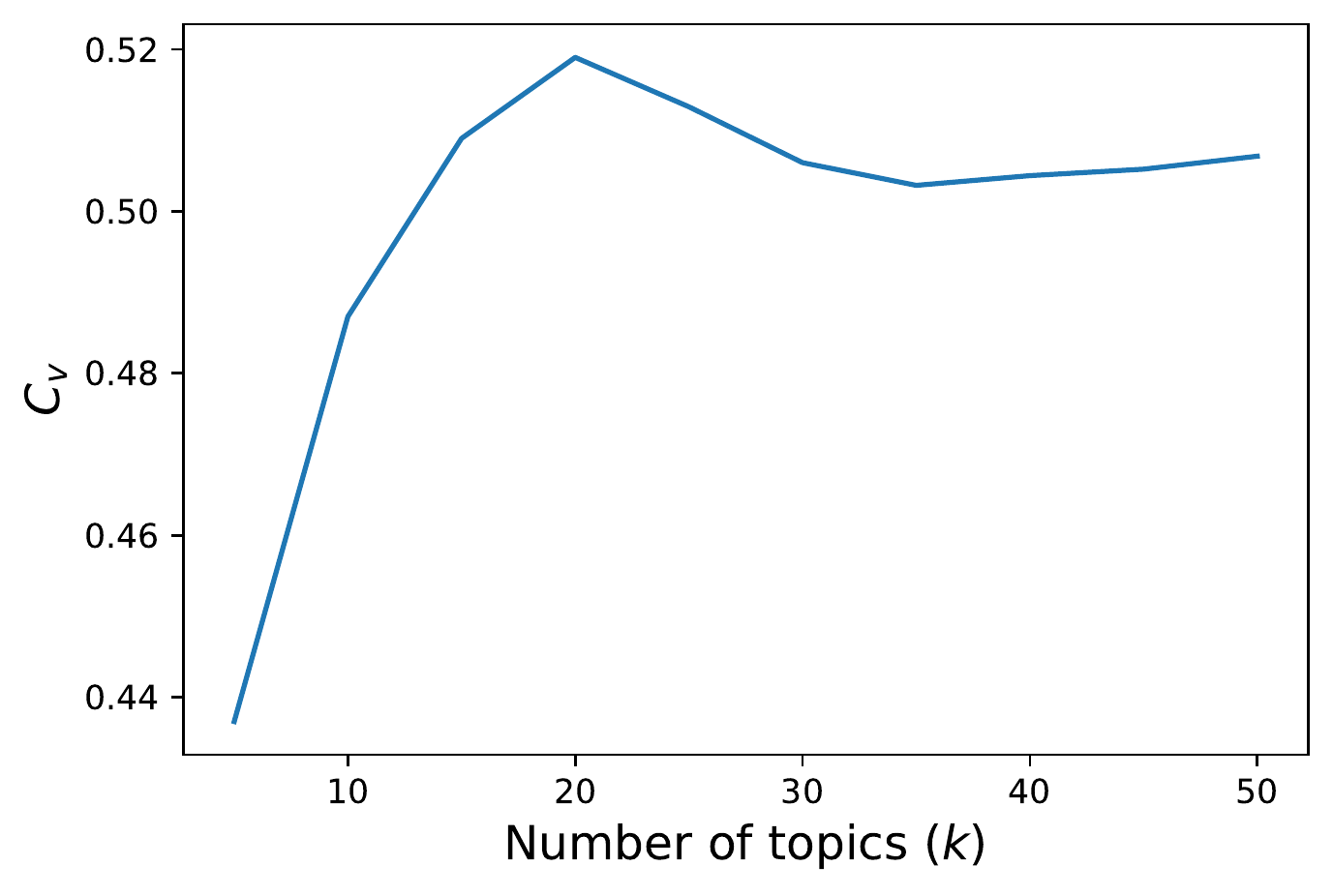}
    \caption{$C_v$ metric according to no of topics.}
    \label{fig:cv_score}
\end{figure}

For training LDA models, we employed the implementation provided by Machine Learning for Language Toolkit (MALLET) \footnote{\url{http://mallet.cs.umass.edu/}}. To obtain the most coherent topic model for our data, we vary the number of topics $k$ from 5 to 50 with a step of 5, and train LDA models with 1,000 Gibbs sampling iterations and priors $\alpha = 5/k$ and $\beta = 0.01$. For each trained model, we compute the $C_v(k)$ metric using the implementation provided by Gensim library \cite{rehurek_lrec}. We find that $k=20$ is the optimal topic number according to the $C_v$ metric ($C_v(20)=0.52$), see Figure \ref{fig:cv_score}.

In Table \ref{tab:topics}, we present the topics learned by our best LDA model, including the most relevant terms describing each topic and the number of chats where each topic is dominant. To obtain the most descriptive terms for topic interpretation, we adopted the approach of ranking individual terms within topics presented in \cite{sievert2014ldavis} and set $\lambda=1$.

\subsection{Topic Interpretation and Analysis}
From Table \ref{tab:topics}, it is evident that the most prevalent topic across all broadcasts in our LDA experiment is topic \#18, dominating the topic mixture proportions in 12,209 chat log documents (19\% of the modelled documents). This topic is clearly related to sexual grooming, with key terms including \textit{CLOTH\_TERM, show, open, SEX\_TERM}, and various other relevant terms previously identified in our grooming behaviour analysis (e.g. remove, wear, top - which in this case refers to a clothing item, etc.).

The second most dominant topic (\#11) reflects flirtatious behaviours, including many endearment terms (e.g. love, nice, pretty, kiss, cute, gorgeous, hot), and words associated with appearance features (e.g. eye, lip, hair, smile, tattoo). Topic \#11 is the most representative of 8,477 chat log documents (13\% of modelled broadcasts). The rest of the topics describe a wide range of behaviours occurring in the context of live streams, including virtual currency and gifts of LiveMe (i.e. coin, coindrop, castle, diamond, wand), dancing, singing, eating, social media, etc. An interesting observation is the emergence of a topic containing mostly Spanish words (topic \#2). We speculate that a proportion of the US viewers are using Spanish to communicate within the broadcasts, something we did not consider in the preprocessing stage. It should be noted that Spanish are the second most spoken language is the US and widely used in some states. Nonetheless, provided that it dominates only 2,142 chats (3\% of modelled broadcasts), we expect that its impact will be negligible for the rest of our analysis.


Next, we assess the degree to which user interactions other than chatting can be characteristic of grooming.
For this, we leverage the interaction and metadata features of our dataset, and additionally, we normalise the interaction features by the total number of viewers of each stream, considering them as additional features. Next, we employ the Mean Decrease Impurity (MDI) \cite{breiman2001random,breiman2002manual} measure obtained in the process of random forest growing to assess the importance of the described features for discriminating between the broadcasts where topic \#18 is dominant in the topic mixture and the rest.

\begin{figure*}[!ht]
  \centering
  \begin{subfigure}[t]{0.32\textwidth}
      \centering
    \includegraphics[width=\textwidth]{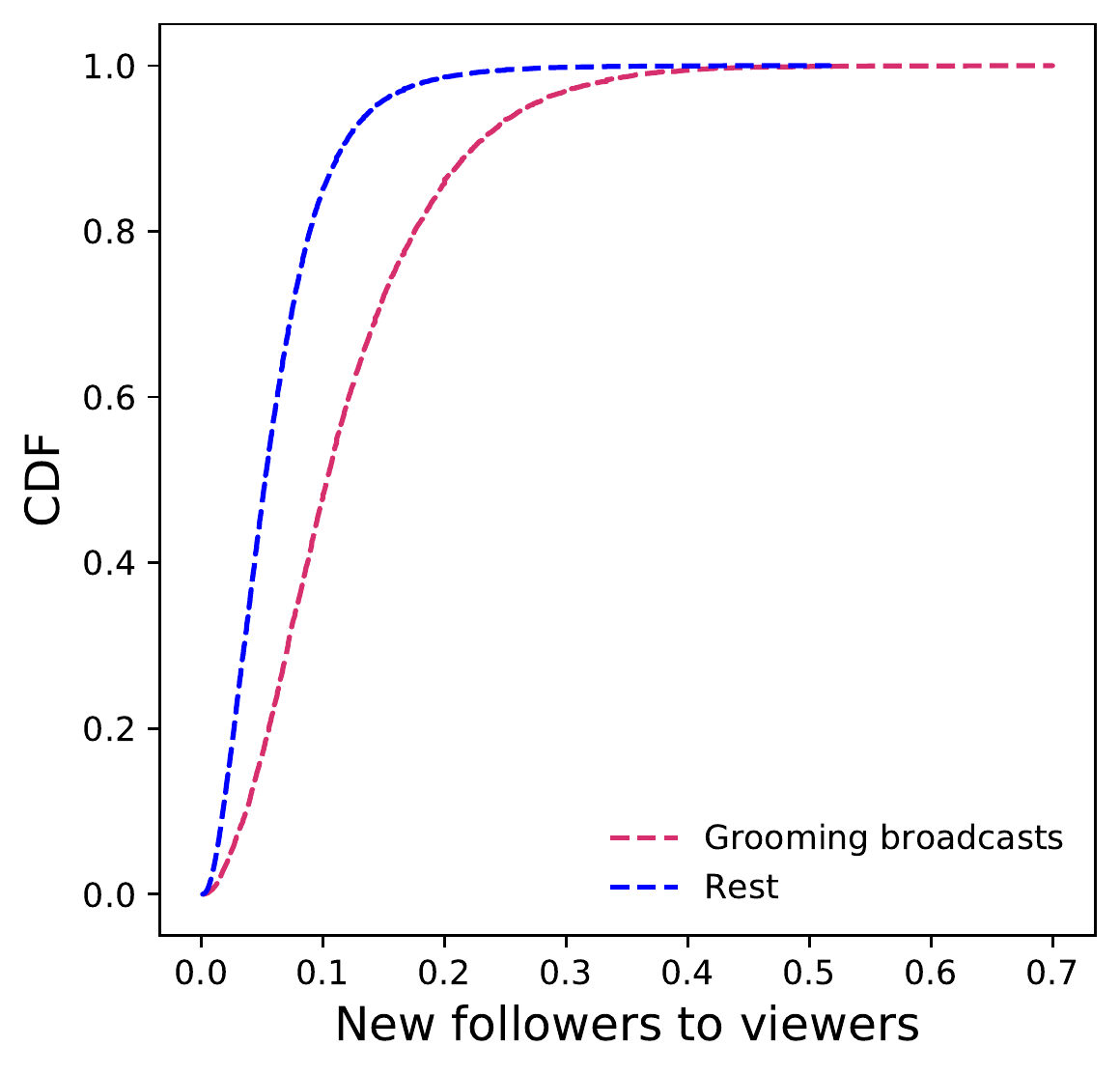}
    \caption{New followers to viewers}
    \label{fig:followers_frac_cdf}
  \end{subfigure}
  \begin{subfigure}[t]{0.32\textwidth}
      \centering
    \includegraphics[width=\textwidth]{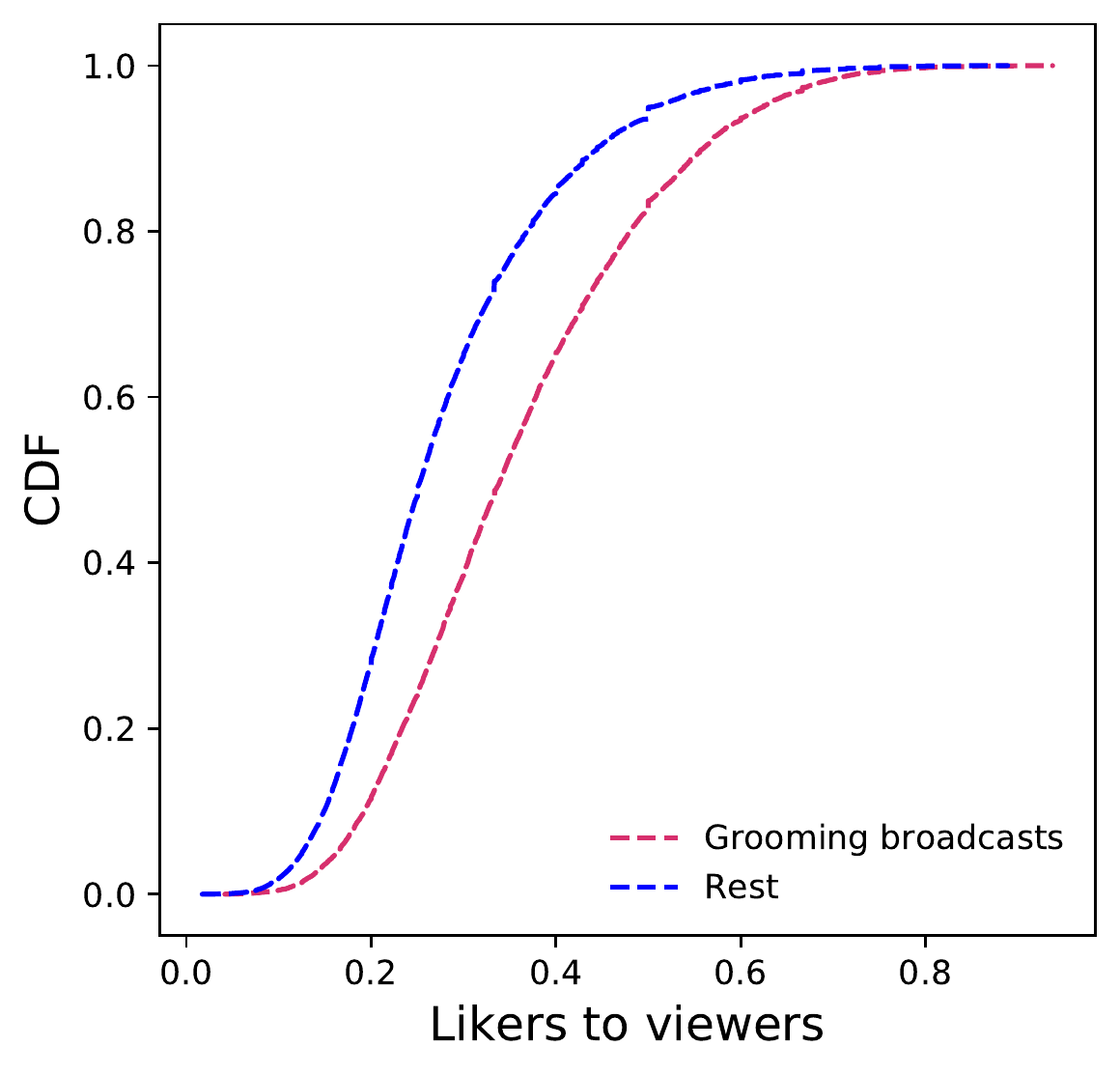}
    \caption{Likers to viewers}
    \label{fig:likers_frac_cdf}
  \end{subfigure}
  \begin{subfigure}[t]{0.32\textwidth}
      \centering
    \includegraphics[width=\textwidth]{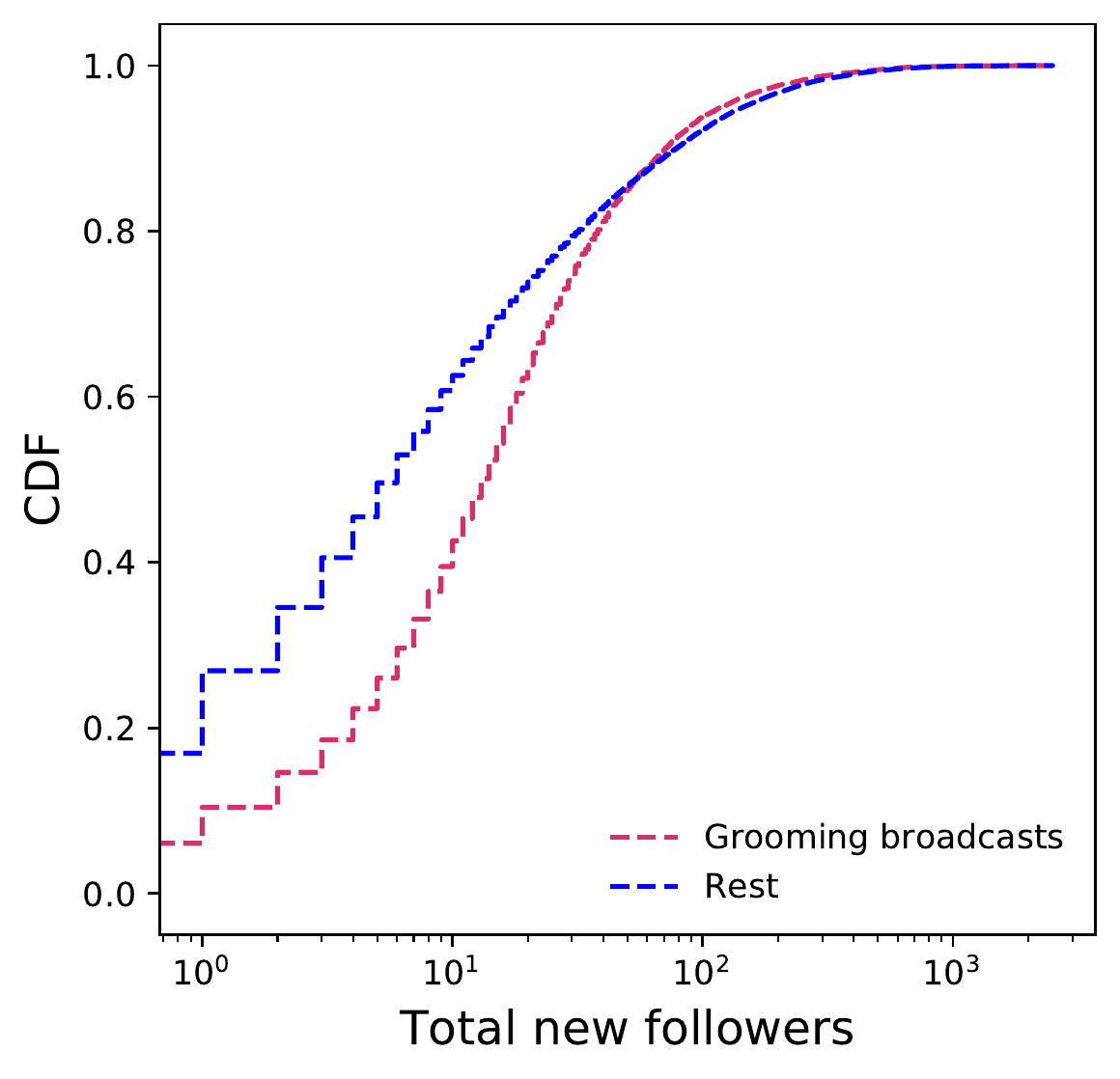}
    \caption{Total new followers}
    \label{fig:followers_cdf}
  \end{subfigure}

  \begin{subfigure}[t]{0.32\textwidth}
      \centering
    \includegraphics[width=\textwidth]{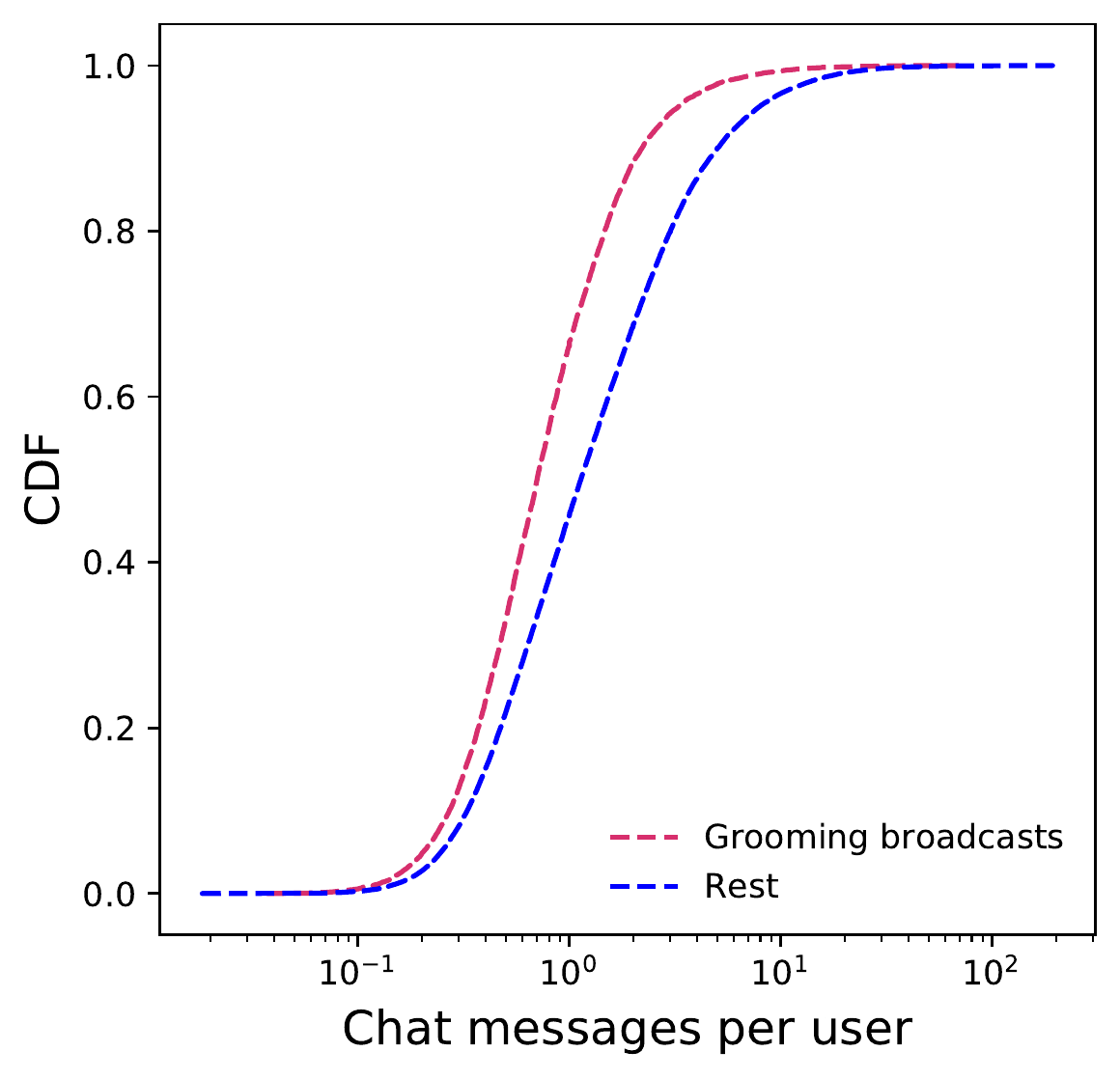}
    \caption{Chat messages per user}
    \label{fig:chat_per_user_cdf}
  \end{subfigure}
  \begin{subfigure}[t]{0.32\textwidth}
      \centering
    \includegraphics[width=\textwidth]{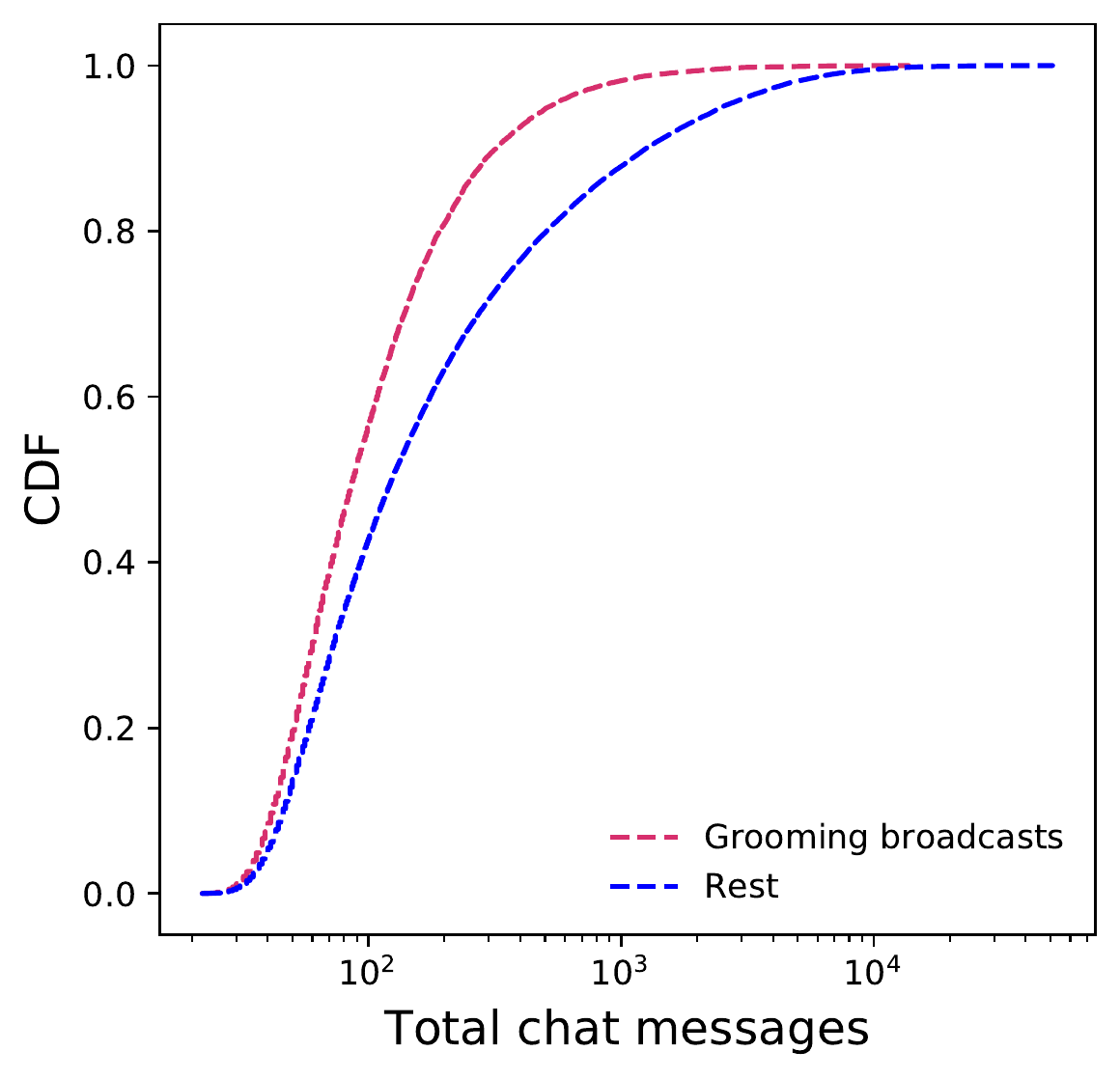}
    \caption{Total chat messages}
    \label{fig:total_chat_cdf}
  \end{subfigure}
  \caption{Cumulative distribution functions (CDFs) of the features of Table \ref{tab:feature-importance}.}
  \label{fig:feature_importance_cdfs}
\end{figure*}

Table \ref{tab:feature-importance} reports the ranking of the top five most important features according to the normalised MDI metric.
To understand how these features are distributed across the two latent classes of broadcasts,  we plot their cumulative distribution functions (CDFs) in Figure \ref{fig:feature_importance_cdfs}.
We note that the most characterising feature is the fraction of viewers who started following the broadcaster during the stream (Fig. \ref{fig:followers_frac_cdf}), which in the case of the broadcasts where the grooming topic dominates is much higher than the ones where it does not. Moreover, in Fig. \ref{fig:followers_cdf} we observe that only around 6\% of the grooming broadcasts have not generated any followers for the broadcaster, while the same is true for 17\% of the rest of broadcasts.
This behaviour is in line with the findings of \cite{lykousas2018adult}, where the adult content producers of LiveMe were found to have a particularly high number of followers which are characterised by their tendency to follow users who have broadcasted adult content systematically, labelled as \textit{adult content consumers}. A possible explanation could be that in broadcasts where the grooming behaviour is prevalent, broadcasters are coerced into performing sexual acts requested by the viewers, as previously outlined. This could explain the reason why the number of new followers they gain in such broadcasts is significantly higher since the viewers might expect that the broadcasters will stream more nude/adult content in the future, and following them is the only way to be notified when they start a new broadcast.
Similarly, the fraction of viewers who have liked a broadcast is higher when the grooming behaviour is dominant (Fig. \ref{fig:likers_frac_cdf}, which is consistent with the findings of \cite{lykousas2018adult} where adult content producers are observed to have received higher amounts of praise than the users found in their ego-networks (i.e. followers and followees). This further exemplifies the predatory behaviour of viewers who use likes/praise to coerce broadcasters into inappropriate acts or reward them when they have achieved their objective.
Interestingly, for the \textit{Chat messages per user} and \textit{Total chat messages} features which were also found to be important (albeit considerably less impactful in a classification setting), we observe the opposite behaviour: In grooming broadcasts users exchange fewer chat messages, both per user and at broadcast level.

\begin{table}
\centering
\rowcolors{2}{gray!25}{white}
\begin{tabular}{llc}
\toprule
\multicolumn{1}{c}{\textbf{Rank}} & \multicolumn{1}{c}{\textbf{Feature}} & \textbf{MDI} \\\midrule
1 & New followers to viewers & 0.36  \\
2 & Likers to viewers & 0.16  \\
3 & Total new followers & 0.10   \\
4 & Chat messages per user & 0.10  \\
5 & Total chat messages & 0.05  \\ \bottomrule
\end{tabular}
\caption{Top 5 interaction features relevant for characterising grooming broadcasts}
    \label{tab:feature-importance}
\end{table}

\begin{sidewaystable}[!th]
    \centering
    \footnotesize
    \rowcolors{2}{gray!25}{white}
\begin{tabular}{rp{6in}r}
\toprule
 \textbf{Topic} & \textbf{Keywords} &  \textbf{\#Docs} \\
\midrule
    18 &  CLOTH\_TERM, show, open, SEX\_TERM, nice, dare, dance, hot, stand, leg, put, kiss, turn, pull, wear, cam, camera, remove, foot, top, snapchat, gift, girl, rub, low, hand, lift, finger, message, tease &           12,209 \\
    11 &  love, nice, pretty, kiss, cute, eye, girl, gorgeous, hot, SEX\_TERM, sweet, lip, hair, dear, tattoo, smile, single, dance, friend, number, stand, beauty, lovely, cutie, face, boyfriend &            8,477 \\
     1 &  sleep, phone, bed, tired, cool, car, wake, cold, drive, smoke, hour, fall, hear, talk,  asleep, stay, high, long, house, game, guess, chill, goodnight, sound, money, iphone, fun, pay  &            5,731 \\
    19 &  talk, hear, happen, friend, leave, wrong, true, cool, mad, sad, care, sound, hurt, smile, fight, stay, fine, dude, person, funny, break, hard, nice, head, long, boy, army, problem, lose, girl, yep &            4,432 \\
     7 &  block, admin, girl, message, leave, report, show, talk, account, kid, creep, young, shut, ban, rude, fake, nasty, send, perv, truth, boy, lie, hater, wrong, police, child, unblock &            3,328 \\
    16 &  drink, cat, food, pizza, laugh, eat, funny, face, water, put, chicken, dead, hair, head, challenge, roast, cream, leave, SEX\_TERM, apple, taco, chocolate, pet, bob, hand, candy, mouth, cheese, nose &            3,180 \\
    12 &  cute, snapchat, send, instagram, rate, clown, dab, hot, number, insta, hair, play, text, friend, pretty, love, put, single, phone, kik, profile, eye, cutie, chat, ghost, boy, girl, fake, girlfriend &            3,080 \\
     3 &  send, gift, spam, castle, diamond, share, top, level, broadcast, win, giveaway, broadcaster, wand, number, stream, boat, enter, entry, star, love, feature, join, porsche, coin, awesome, comment, fan &            2,953 \\
     5 &  coin, drop, coindrop, follower, send, win, feature, shout, fan, castle, love, wand, dab, thot, number, gift, shoutout, giveaway, diamond, stream, iphone, lag, goal, dude, pumpkin, andy, light, level &            2,798 \\
     4 &  song, play, sing, love, voice, rap, singing, amazing, nice, dance, awesome, beat, put, hear, panda, listen, cool, singer, sound, closer, juju, black, job, girl, talent, guitar, boy, heart, hit, drake &            2,687 \\
     8 &  love, stream, friend, accent, talk, remember, guess, cool, speak, leave, sleep, skype, long, cute, funny, nice, number, meet, hair, mate, lot, person, dad, class, cat, joke, jenni, kat, join, change &            2,682 \\
    20 &  light, turn, gang, love, stay, queen, squad, hit, chill, number, king, slay, fact, thot, level, savage, rock, party, dead, boy, mad, play, homie, ight, lot, black, nun, show, petty, dope, top, sum &            2,366 \\
     2 &  hola, mami, como, cute, hermosa, eres, show, spanish, amor, SEX\_TERM, pretty, bella, donde, hot, bonita, bien, lip, kiss, speak, tienes, espanol, gorgeous, stand, rico, jada &            2,142 \\
    13 &  girl, love, cute, play, blue, twin, pretty, red, hot, black, dance, snapchat, green, pink, makeup, hair, lady, white, friend, cool, color, game, face, team, texas, nice, CLOTH\_TERM, batman, favorite &            1,954 \\
    14 &  kate, love, kid, nice, awesome, cool, tree, country, santa, dad, boy, level, show, broadcast, send, amazing, hear, wolf, talk, lot, son, king, falcon, grim, happen, stream, matt, house, long, rock &            1,831 \\
     9 &  beam, love, lag, send, king, cris, stream, castle, fletch, broadcast, show, level, dude, awesome, nick, game, amazing, feature, remember, joey, gift, beem, roll, diamond, join, happen, rip, rackbar &            1,663 \\
    15 &  ready, love, spam, feature, stay, game, number, win, boy, tre, read, letter, chat, duck, turtle, greg, cat, spamme, fun, red, ugh, play, controller, send, coin, hehe, cool, high, comment, gift, party &            1,027 \\
    17 &  love, fan, favorite, youtube, shout, meet, dab, channel, pickle, song, shoutout, canada, movie, awesome, fav, twerk, vote, magic, food, subscribe, notice, tattoo, cool, texas, win, vid, hair, ily &             908 \\
     6 &  race, love, family, human, unity, amen, put, country, draw, earth, whiskey, broadcast, peace, block, thre, lucky, princess, spam, britt, join, general, respect, coin, barbie, send, level, lag, brit &             642 \\
    10 &  president, kira, criticize, article, essay, literary, loco, fard, natur, fward, lag, foard, riot, ward, folard, kilo, follrd &              14 \\
\bottomrule
\end{tabular}
\caption{Topics}
\label{tab:topics}
\end{sidewaystable}

\subsection{Topic relatedness}
In this section, we aim to explore the relatedness of the dominant grooming topic and other topics learned by LDA, which could unveil different aspects of this deviant behaviour, beyond our initial analysis.  To this end, we use a frequent itemset mining approach to examine the co-occurrence of prevalent topics within the chat log documents. More precisely, we first selected the three topics with the highest probability in the mixture assigned to each broadcast, and then applied the FP-growth algorithm \cite{han2004mining} to discover frequent patterns of size two. In table \ref{tab:fp-growth}, we show the 10 most frequent patterns extracted following the described approach. As expected, the top result includes the two most prevalent topics in the mixture. Notably, the second most frequent pattern includes the grooming topic, and topic \#7, which contains terms related to the (self) moderation of broadcasts (i.e. block, report, ban, shut), terms indicating young age (i.e. kid, young, child, girl, boy), terms of hostility (i.e. creep, perv, hater), words bearing negative sentiment according to LIWC (nasty, wrong, fake, lie). Moreover, the key term that possibly contributes the most towards the interpretation of this topic is \textbf{police}. Thus, we expect this topic to be indicative of the criminal dimension of sexual grooming of minors in LiveMe, a large-scale deviant behaviour, also attested by popular media \cite{fox}.

To test this speculation, we manually examined a portion of chat messages from broadcasts where Topic \#7 is dominant where the aforementioned key terms appear, and we present some illustrative examples in Table \ref{tab:sample}. What we observe is that a part of the users expresses their discontent and anger towards the predators/groomers and their harassment targeting minors. We argue that the above illustrates the extent of deviant behaviour in SLSS, something that beyond the media is also reported by users in, e.g. their feedback for the app. Moreover, the high ranking of this pair indicates that such phenomena, despite the app's moderation mechanisms, are often and known to many users. Finally, the fourth pair (13,18) beyond the common keywords of both topics, shows that some users request further engagement through other platforms, and a primary phase of praise of clothing and body parts, possibly preceding the grooming phase.

\begin{table}
\centering
    \rowcolors{2}{gray!25}{white}
    \begin{tabular}{p{0.95\columnwidth}}
    \toprule
        {\bf Chat message} \\
        \midrule
        A predator is a person who asks \red{\textbf{kids}} to undress in front of the camara \\
        And his bio said he likes meeting \red{\textbf{young} \textbf{girls}} \\
        \red{\textbf{Block}} foot fetish \red{\textbf{creep}} \\
        Don't show the \red{\textbf{creeps}} anything \\
        Everyone \red{\textbf{report}} chat \red{\textbf{police}} \\
        Leave her alone \red{\textbf{creep}} \\
        \red{\textbf{Pervs}}. This \red{\textbf{kid}} is like 12 \\
        \red{\textbf{Report}} that \red{\textbf{creep}} too the \red{\textbf{police}} \\
        \red{\textbf{Report}} the users asking \red{\textbf{kids}} to undress; to authorities not LiveMe \\
        Show your \red{\textbf{kids}} \\
        So if they didn't \red{\textbf{ban}} people for nudity you would show? \\
        This needs to be \red{\textbf{reported}} what sort of sick people are ye. She is only 11 \\
        YOUR MOM WILL NOW GET A CALL TO KNOW YOU TALK TO 40 years old \red{\textbf{creeps}} \\
        You \red{\textbf{pervs}} are \red{\textbf{nasty}} as f*** \\
        \red{\textbf{block}} \& \red{\textbf{report}} \red{\textbf{nasty}} stuff \\
        \red{\textbf{creeps}} make \red{\textbf{kids}} do \red{\textbf{nasty}} stuff \\
        he's following lots of \red{\textbf{young} \textbf{girls}} \\
        \red{\textbf{pervs}} stop asking her to undress \\
        \red{\textbf{report}} these pedos to \red{\textbf{police}} mate \\
        she is a \red{\textbf{child}} stop asking that \\
        she not leting you \red{\textbf{creeps}} or sick \red{\textbf{perv}} seeing her dress or undress ok \\
        she's a \red{\textbf{kid}}. ...\red{\textbf{perv}} \\
        they can't \red{\textbf{ban}} you if you delete your video after you show \\
        too \red{\textbf{young}} this is illegal nd worng lol \\
        try not to undress on stream, it will draw in a lot of \red{\textbf{creeps}} \\
        you have \red{\textbf{creeps}} who made you do \red{\textbf{nasty}} stuff \\
        you look very \red{\textbf{young}}.there are lots of pedos on here. be careful \\
    \bottomrule
    \end{tabular}
    \caption{Illustrative chat messages from broadcasts where Topic \#7 is dominant. Key terms of Topic \#7 are in bold.}
    \label{tab:sample}
\end{table}


\begin{table}
\rowcolors{2}{gray!25}{white}
\centering
\caption{10 most frequent prevalent-topic patterns}
    \label{tab:fp-growth}
\begin{tabular}{lr}
\toprule
\textbf{Topic pattern} &  \textbf{Occurrences} \\
\midrule
     (11, 18) &        11,150 \\
      (7, 18) &         5,536 \\
      (1, 19) &         4,892 \\
     (13, 18) &         3,843 \\
      (1, 16) &         3,416 \\
      (1, 11) &         3,357 \\
      (7, 11) &         3,338 \\
       (3, 5) &         3,331 \\
      (2, 11) &         3,269 \\
     (12, 18) &         2,987 \\
\bottomrule
\end{tabular}
\end{table}

\section{Conclusions}
Social live streaming services due to the continuous use of live streams and immediate user interaction are continuously expanding their user base. As expected, these platforms have attracted the interest of deviant users which try to exploit the new features on these platforms. Obviously, grooming is not only performed in SLSS, nor it is the only thing done on these platforms. Nonetheless, the different possible user interactions coupled with the live streaming nature create a novel and less explored field.

This work performs an in-depth analysis of the chats of thousands of users and identifies characteristics of the grooming behaviour in the verbal and non-verbal context. To facilitate further research in the field, we responsibly share a massive dataset and provide ethical and legal justification for the collection and processing of such a dataset. Moreover, we illustrate in an automated way how users bypass the word filters that service providers use in their platforms. We also highlight the importance of emojis for the first time in the context of grooming. Finally, our work illustrates that more deviant behaviours may be performed on these platforms.

We believe that this scientific work constitutes a significant contribution towards understanding the deviant behaviours on social networks. The latter should be considered in the light of the role that social networks have in our daily lives and the potentials that the emergence of SLSS have. Our work implies that further risks exist due to the inefficiency of current moderation mechanisms. Therefore, further measures must be taken to secure the content of what is broadcast, from whom, and to whom. Undoubtedly, due to the size and rate of exchanged information, moderation mechanisms may be difficult to be performed in real-time. However, our work illustrates how deviant behaviours can be detected effectively without resolving to the use of multimedia which require heavy processing. Therefore, we believe that the grooming and other predatory actions will be soon identified better and addressed more effectively by service providers.

\section*{Acknowledgements}
This work was supported by the European Commission under the Horizon 2020 Programme (H2020), as part of the projects CyberSec4Europe (\url{https://www.cybersec4europe.eu}) (Grant Agreement no. 830929) and \textit{LOCARD} (\url{https://locard.eu}) (Grant Agreement no. 832735). The authors would also like to thank NVIDIA Corporation for their GPU donation supporting their research.

The content of this article does not reflect the official opinion of the European Union. Responsibility for the information and views expressed therein lies entirely with the authors.

\end{document}